\documentclass[12pt]{article}
\usepackage{epsfig}
\usepackage{amsmath}
\usepackage{hhline}
\usepackage{amssymb}
\usepackage{times}
\usepackage{cite}

\newlength{\dinwidth}
\newlength{\dinmargin}
\begin{document}
\newcommand {\gapprox}
   {\raisebox{-0.7ex}{$\stackrel {\textstyle>}{\sim}$}}
\newcommand {\lapprox}
   {\raisebox{-0.7ex}{$\stackrel {\textstyle<}{\sim}$}}
\def\gsim{\,\lower.25ex\hbox{$\scriptstyle\sim$}\kern-1.30ex%
\raise 0.55ex\hbox{$\scriptstyle >$}\,}
\def\lsim{\,\lower.25ex\hbox{$\scriptstyle\sim$}\kern-1.30ex%
\raise 0.55ex\hbox{$\scriptstyle <$}\,}
\newcommand{\gap}{\stackrel{>}{\sim}}
\newcommand{\lap}{\stackrel{<}{\sim}}
\newcommand{\delstat}{$\delta_{\mathrm{stat}}$}
\newcommand{\delplus}{$\delta_{\mathrm{syst}}$}
\newcommand{\delminus}{hadr. corr.}
\newcommand{\qiaa}{2.0 -- 4.4}
\newcommand{\qibb}{4.4 -- 10}
\newcommand{\qicc}{10 -- 25}
\newcommand{\qidd}{25 -- 80}
\newcommand{\eiaa}{5 -- 10}
\newcommand{\eibb}{10 -- 20}
\newcommand{\eicc}{20 -- 60}
\newcommand{\eyaa}{5 -- 7}
\newcommand{\eybb}{7 -- 10}
\newcommand{\eycc}{10 -- 15}
\newcommand{\eydd}{15 -- 20}
\newcommand{\eyee}{20 -- 30}
\newcommand{\eyff}{30 -- 60}
\newcommand{\xiaa}{0.12 -- 0.35}
\newcommand{\xibb}{0.35 -- 0.55}
\newcommand{\xicc}{0.55 -- 0.75}
\newcommand{\xidd}{0.75 -- 1.00}
\newcommand{\xyaa}{0 -- 0.75}
\newcommand{\xybb}{0.75 -- 1}
\newcommand{\yiaa}{0.10 -- 0.25}
\newcommand{\yibb}{0.25 -- 0.50}
\newcommand{\yicc}{0.50 -- 0.85}
\newcommand{\yyaa}{0.10 -- 0.25}
\newcommand{\yybb}{0.25 -- 0.40}
\newcommand{\yycc}{0.40 -- 0.55}
\newcommand{\yydd}{0.55 -- 0.70}
\newcommand{\yyee}{0.70 -- 0.85}
\newcommand{\riaa}{-2.5 -- (-2.0)}
\newcommand{\ribb}{-2.0 -- (-1.5)}
\newcommand{\ricc}{-1.5 -- (-1.0)}
\newcommand{\ridd}{-1.0 -- (-0.5)}
\newcommand{\riee}{-0.5 -- 0.0}
\newcommand{\ryaa}{-2.5 -- (-1.7)}
\newcommand{\rybb}{-1.7 -- (-1.3)}
\newcommand{\rycc}{-1.3 -- 0}
%\newcommand{\GeV}{\,\mbox{GeV}}
%\newcommand{\MeV}{\,\mbox{MeV}}
%
% Some useful tex commands
%
\newcommand{\gammaT}{\ensuremath{\gamma^*_T} }
\newcommand{\gammaL}{\ensuremath{\gamma^*_L} }
\newcommand{\xg}{\ensuremath{x_\gamma} }
\newcommand{\xgmy}{\ensuremath{x^{\mathrm{jets}}_\gamma} }
\newcommand{\qq}{\ensuremath{Q^2} }
\newcommand{\GeV}{\ensuremath{\mathrm{GeV}} }
\newcommand{\pb}{\ensuremath{\mathrm{pb}} }
\newcommand{\gevsq}{\ensuremath{\mathrm{GeV}^2} }
\newcommand{\Etone}{\ensuremath{E^*_{T\, 1}} }
\newcommand{\Ettwo}{\ensuremath{E^*_{T\, 2}} }
\newcommand{\Etmy}{\ensuremath{E^*_T} }
\newcommand{\Etsq}{\ensuremath{E_T^2} }
\newcommand{\Etmysq}{\ensuremath{E_T^{*\, 2}} }
\newcommand{\etaone}{\ensuremath{\eta^*_{1}} }
\newcommand{\etatwo}{\ensuremath{\eta^*_{2}} }
\newcommand{\etamy}{\ensuremath{\eta^*} }
\newcommand{\gp}{\ensuremath{\gamma^*}p }
\newcommand{\ycut}{\ensuremath{y_c} }
\newcommand{\PDFgamma}{\ensuremath{D_{i/\gamma^*}} }
\newcommand{\PDFgammaQED}{\ensuremath{D_{i/\gamma^*}^{\mathrm{QED}}} }
\newcommand{\PDFgammaT}{\ensuremath{D_{i/\gamma^*_T}} }
\newcommand{\PDFgammaL}{\ensuremath{D_{i/\gamma^*_L}} }
\newcommand{\PDFproton}{\ensuremath{D_{j/p}} }
\newcommand{\lumi}{57~pb$^{-1}$}
\newcommand{\lambdaQCD}{\ensuremath{\Lambda_{\mathrm{QCD}}} }
\newcommand{\muf}{\ensuremath{\mu_f}}
\newcommand{\mufsq}{\ensuremath{\mu_f^2}}
\newcommand{\mur}{\ensuremath{\mu_r}}
\newcommand{\mursq}{\ensuremath{\mu_r^2}}
\newcommand{\muff}{\ensuremath{\hat{\mu}_f}}
\newcommand{\muffsq}{\ensuremath{\hat{\mu}^2_f}}
\newcommand{\sigone}{$\frac{{\rm d}^3\sigma_{\rm {2jet}}}{{\rm d}Q^2 {\rm d}\Etmy\, {\rm d}\xgmy}(\frac{\pb}{\GeV^3})$}
\newcommand{\sigtwo}{$\frac{{\rm d}^3\sigma_{\rm {2jet}}}{{\rm d}Q^2 {\rm d}y {\rm d}\etamy}(\frac{\pb}{\GeV^2})$}
\newcommand{\sigthree}{$\frac{{\rm d}^3\sigma_{\rm {2jet}}}{{\rm d}Q^2 {\rm d}\etamy\, {\rm d}\Etmy}(\frac{\pb}{\GeV^3})$}
\newcommand{\sigfour}{$\frac{{\rm d}^3\sigma_{\rm {ep}}}{{\rm d}Q^2 {\rm d}\xgmy {\rm d}y}(\frac{\pb}{\GeV^2})$}
\newcommand{\dsqex}{${\rm d}^3\sigma_{\rm {2jet}}/{\rm d}Q^2 {\rm d}\Etmy\, {\rm d}\xgmy$ }
\newcommand{\dsqyr}{${\rm d}^3\sigma_{\rm {2jet}}/{\rm d}Q^2 {\rm d}y {\rm d}\etamy$ }
\newcommand{\dsqre}{${\rm d}^3\sigma_{\rm {2jet}}/{\rm d}Q^2 {\rm d}\etamy\, {\rm d}\Etmy$ }
\newcommand{\dsqxy}{${\rm d}^3\sigma_{\rm {ep}}/{\rm d}Q^2 {\rm d}\xgmy {\rm d}y$ }
\newcommand{\dsqy}{${\rm d}^2\sigma_{\rm {ep}}/{\rm d}Q^2 {\rm d}y$ }
%
% Journal macro
\def\Journal#1#2#3#4{{#1} {\bf #2} (#3) #4}
\def\NCA{\em Nuovo Cimento}
\def\NIM{\em Nucl. Instrum. Methods}
\def\NIMA{{\em Nucl. Instrum. Methods} {\bf A}}
\def\NPB{{\em Nucl. Phys.}   {\bf B}}
\def\PLB{{\em Phys. Lett.}   {\bf B}}
\def\PRL{\em Phys. Rev. Lett.}
\def\PRD{{\em Phys. Rev.}    {\bf D}}
\def\ZPC{{\em Z. Phys.}      {\bf C}}
\def\EJC{{\em Eur. Phys. J.} {\bf C}}
\def\CPC{\em Comp. Phys. Commun.}

\hyphenation{ca-lo-ri-me-ter}

\begin{titlepage}

\begin{figure}[!t]
DESY--03--206 \hfill ISSN 0418--9833\\ 
%\hfill hep--ph/03xxxxx\\
January 2004
\end{figure}
\bigskip

\vspace*{2cm}

\begin{center}
\begin{Large}

{\bf Measurement of Dijet Production at Low
    {\boldmath $Q^2$} at HERA}

\vspace{2cm}

H1 Collaboration

\end{Large}
\end{center}

\vspace{2cm}

\begin{abstract}
Triple differential dijet cross sections in $e^\pm p$ interactions
are presented in the region
of photon virtualities $2 < Q^2 < 80\, \gevsq$, inelasticities
$0.1<y<0.85$, jet transverse energies $\Etone>7\, \GeV$, $\Ettwo>5\,
\GeV$, and pseudorapidities $-2.5 < \etaone, \etatwo <0$.
The measurements are made in the $\gamma^* p$ 
centre-of-mass frame, using an integrated
luminosity of \lumi. The data are
compared with NLO QCD calculations and LO Monte Carlo
programs with and without a resolved virtual photon contribution.
NLO QCD calculations fail to describe the region of low $Q^2$
and low jet transverse energies, in contrast to a LO Monte
Carlo generator
which includes direct and resolved photon interactions with both
transversely and longitudinally polarised photons.
Initial
and final state parton showers are tested as a mechanism for including
higher order QCD effects in low $E_T$ jet production.
\end{abstract}

\vspace{1.5cm}

\begin{center}
To be submitted to Eur.\ Phys.\ J.\ C
\end{center}

\end{titlepage}

\begin{flushleft}
  %-- H1AUTS Author list by names 
%-- Status: Mon May 12 10:17:47 MET DST 2003  Number of authors = 311 

A.~Aktas$^{10}$,               %DESY-ST        03/2            Aktas               
V.~Andreev$^{24}$,             %LPI -PD        8/88            Andreev             
T.~Anthonis$^{4}$,             %ANTW-ST        11/99           Anthonis            
A.~Asmone$^{31}$,              %ROME-ST        07/2            Asmone              
A.~Babaev$^{23}$,              %ITEP-PD        8/88            Babaev              
S.~Backovic$^{35}$,            %ZEUT-PD        03/2            Backovic            
J.~B\"ahr$^{35}$,              %ZEUT-PD        8/88            Baehr               
P.~Baranov$^{24}$,             %LPI -PD        8/88            Baranovp            
E.~Barrelet$^{28}$,            %PARI-PD        11/99           Barrelet            
W.~Bartel$^{10}$
% ...

%-- H1 Institutes 
\bigskip{\it
% ...
$ ^{4}$ Inter-University Institute for High Energies ULB-VUB, Brussels;
          Universiteit Antwerpen (UIA), Antwerpen; Belgium$^{ c}$ \\
% ...
$ ^{10}$ DESY, Hamburg, Germany \\
% ...
$ ^{23}$ Institute for Theoretical and Experimental Physics,
          Moscow, Russia$^{ l}$ \\
 $ ^{24}$ Lebedev Physical Institute, Moscow, Russia$^{ e}$ \\
% ...
$ ^{28}$ LPNHE, Universit\'{e}s Paris VI and VII, IN2P3-CNRS,
          Paris, France \\
% ...
$ ^{31}$ Dipartimento di Fisica Universit\`a di Roma Tre
          and INFN Roma~3, Roma, Italy \\
% ...
$ ^{35}$ DESY, Zeuthen, Germany \\
% ...
$ ^c$ Supported by FNRS-FWO-Vlaanderen, IISN-IIKW and IWT \\
% $ ^d$ Partially Supported by the Polish State Committee for Scientific
%      Research, grant no. 2P0310318 and SPUB/DESY/P03/DZ-1/99
%      and by the German Bundesministerium f\"ur Bildung und Forschung \\
% ...
$ ^e$ Supported by the Deutsche Forschungsgemeinschaft \\
% ...
$ ^l$ Partially Supported by Russian Foundation
      for Basic Research, grant    no. 00-15-96584 \\
% $ ^m$ Deceased \\
}

\end{flushleft}

\newpage
%\linenumbers
%%%%%%%%%%%%%%%%%%%%%%%%%%%%%%%%%%%%%%%%%%%%%%%%%%%%%%%
%
%
%\section{Introduction}
\section{Introduction}
\label{intro}
Jet cross sections in electron-proton collisions are
successfully described by next-to-leading order
(NLO) QCD calculations in most of the HERA kinematic 
range~\cite{agreementH1,agreementZEUS,Wobisch,Zeus4,Poeschl,Forward}.
However, regions of phase space have previously been 
observed for which NLO predictions do not reproduce the data 
satisfactorily and leading order (LO) Monte
Carlo simulations with different approaches to modelling higher order QCD 
effects are often more 
successful~\cite{Wobisch,Breitweg:2000sv,Zeus4,Poeschl,Forward,Zeus9}.

At HERA, a photon coupling to the incoming electron
interacts with a parton from the proton. The measurement of dijet
production is particularly suitable for the investigation of effects
related to photon structure,
which have been studied
during the last two decades in $e^+e^-$ and $ep$ 
collisions~\cite{expgama,theogama}.
In the ``photoproduction'' region of $ep$ interactions, 
i.e. for $Q^2\ll \Lambda_{\mathrm{QCD}}^2$,
the interaction can be described by the sum of two contributions.
In the {\em direct photon} process, the photon interacts as a whole
with a parton from the proton, whereas in the {\em resolved photon}
process, it behaves as a source of partons, which 
interact with partons from the proton.

As the virtuality of the photon increases, the role
of photon structure gradually changes. Whereas for
quasi-real photons
it is an indispensable theoretical tool, for $Q^2
\gg \Lambda^2_{\mathrm{QCD}}$ the concept of a resolved photon is
usually discarded and the data are analysed within the framework
of perturbative calculations of direct photon processes. However,
it has been argued \cite{sas,GRV,KP,my} that the concept of 
the resolved photon is very useful phenomenologically for arbitrary $Q^2$,
provided the photon virtuality remains much smaller than some
measure of the hardness of the process in which the photon
participates. In our case this is satisfied for $Q^2\ll E_T^2$, where $E_T$
denotes the jet transverse energy. Experimental evidence for the
resolved virtual photon contribution has been found in a
number of publications
\cite{Tanja,Breitweg:2000sv,Nekdo}.

In this paper, new data on dijet production, obtained with the H1
detector in the kinematic region of low to moderate photon virtualities 
$2 < Q^2 < 80\,\GeV^2$, are presented 
as triple differential distributions in $Q^2$, the inelasticity
$y$ and variables characterising the final state jets.
The data are compared with predictions
within several theoretical approaches, differing in the way
QCD effects are taken into account beyond LO, in order to
identify which of them are successful in which regions. 
In doing that we use NLO calculations (i.e. including terms up to 
order $\alpha\alpha_s^2$)
as well as LO calculations supplemented with parton showers, 
which take into account leading logarithmic contributions 
to all orders. The effects of resolved virtual photons are studied for
both transverse and longitudinal photon polarisations.

The paper 
is organised as follows. 
After a review of various theoretical approaches to 
the description of interactions of virtual photons in
Section~\ref{theory}, a brief description of the detector is given
in Section~\ref{detector}. The data sample and event selection are
specified in Section~\ref{cuts}. Background subtractions, 
detector corrections and estimates of the measurement
uncertainties are discussed in
Section~\ref{systematics}. The results are presented and discussed
in Section~\ref{results}.

%%%%%%%%%%%%%%%%%%%%%%%%%%%%%%%%%%%%%%%%%%%%%%%%%%%%%%%
%
%
\section{Dynamics of Hard Processes in {\boldmath $ep$} Collisions}
\label{theory}
\subsection{The DGLAP Approach}
\label{DGLAP_approach}
The DGLAP approach uses parton distribution functions
(PDFs) of the proton 
extracted from
analyses of data on hard scattering processes. These PDFs depend on the
factorisation scale~\muf,
satisfy the DGLAP~\cite{DGLAP} evolution equations,
and are sometimes called ``integrated'' to
emphasise the fact that their definition involves an integral over the
virtualities of partons up to~\mufsq.

\begin{figure}\centering
\epsfig{file=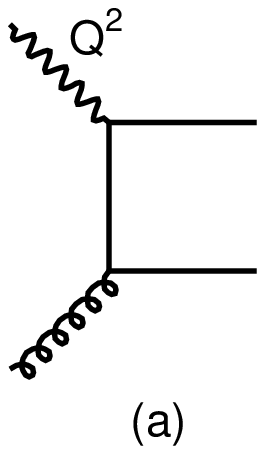,height=5cm}
\epsfig{file=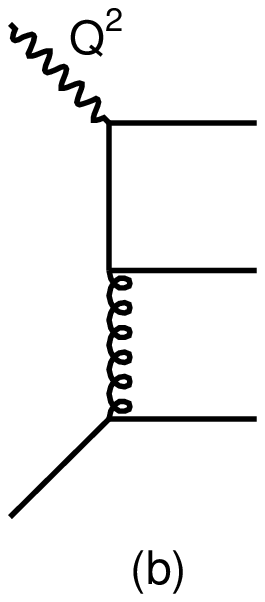,height=5cm}
\epsfig{file=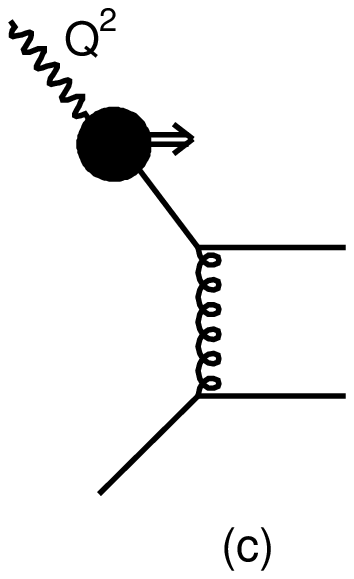,height=5cm}
\epsfig{file=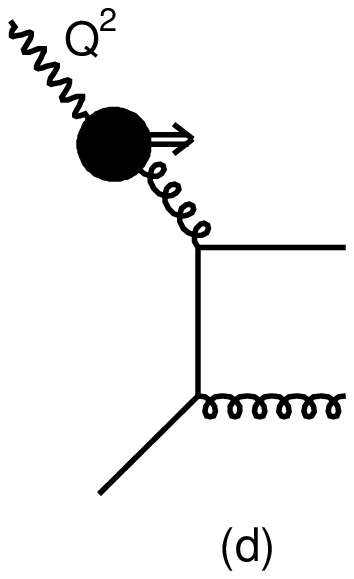,height=5cm}
\epsfig{file=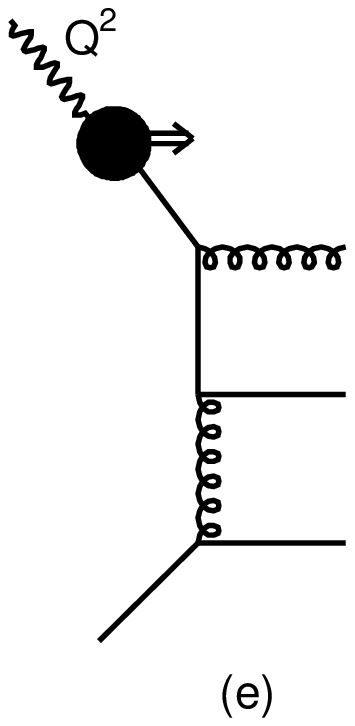,height=5cm}
\caption{Examples of diagrams for the production of at least two jets
(incoming electrons and protons not shown):
  LO  {\em (a)} and NLO {\em (b)} direct photon interactions;
  LO resolved photon interactions involving a quark {\em (c)}
  or a gluon {\em (d)} from the photon; NLO resolved photon
  process~{\em (e)}.}
\label{Feynman}
\end{figure}
%In our kinematic region
%the dominant part of 
In this approach,
the cross section for dijet production in our kinematic region
is given by the direct photon
contribution, illustrated in Fig.~\ref{Feynman}\,{\em{a,b}},
and expressed as
\begin{equation}
    \sigma^{\rm DIR}  \sim
               \sum\limits_j
          D_{j/p} \otimes
         \sigma_{ej} \ ,
\hspace{1cm}
\sigma_{ej}=c^{(1)}_{ej}\alpha_s+c^{(2)}_{ej}\alpha_s^2+\cdots \ ,
\label{direq}
\end{equation}
where $\sigma_{ej}$ denotes the cross section for a collision between 
the incoming electron and a parton~$j$ from the proton, 
$c^{(1)}_{ej}, c^{(2)}_{ej}, \cdots$ 
are coefficients of an expansion of $\sigma_{ej}$ in powers of 
$\alpha_s$ and $D_{j/p}$ denotes the PDF of the proton.
The term $c^{(1)}_{ej}\alpha_s$ defines the LO 
cross section, whilst $~c^{(1)}_{ej}\alpha_s+c^{(2)}_{ej}\alpha_s^2~$ 
defines the NLO cross section.

Recent analyses~\cite{Tanja,Breitweg:2000sv,Poeschl} of dijet
cross sections in the region $\Lambda^2_{\mathrm{QCD}} < Q^2 < E_T^2$ have convincingly shown
that the LO direct photon contribution lies
significantly below the data. The NLO
calculations, involving diagrams such as that shown 
in Fig.~\ref{Feynman}\,{\em b},
bring the theoretical prediction closer to the
data~\cite{Poeschl}.
A recent H1 analysis~\cite{Forward} indicates that even the NLO
calculations do not completely describe inclusive jet
production at low $Q^2$ in part of the phase space.
Large values of the 
NLO corrections, i.e. the ratio of NLO to LO
predictions for the cross sections, and high sensitivity of the predicted jet
cross sections
to variations of the factorisation and renormalisation scales,
strongly suggest the need for higher order 
(i.e. $c^{(3)}_{ej}, c^{(4)}_{ej}, \cdots$) terms in Eq.~(\ref{direq}).
In the absence of a full calculation beyond NLO, some approximate
procedure for resummation of the dominant higher
order terms in Eq.~(\ref{direq}) can be constructed.
This procedure is based on 
the fact that in part of the phase space, the upper vertex of the
diagram in Fig.~\ref{Feynman}\,{\em b} can be
viewed as the splitting of the photon into a $q\overline{q}$ pair.
Taking into account subsequent emissions of partons from this
$q\overline{q}$ pair,
these terms can be resummed into the PDF of the 
photon\footnote{This resummation actually yields the point-like 
(sometimes called ``anomalous'') parts of the photon PDF, 
which dominate in the kinematic region studied in this paper,
where the hadronic (sometimes called ``VMD'') 
parts of the photon PDF are negligible.}, 
$D_{i/\gammaT}$,
as is done for instance in~\cite{sas},
and $D_{i/\gammaL}$ in~\cite{gammal},
where $\gammaT$ denotes the transversely and $\gammaL$ 
the longitudinally polarised virtual photon~\cite{Friberg:2000nx,smarkem1}.

Consequently, one can calculate the resolved photon contribution 
to the dijet cross section, 
corresponding to the graphs shown in Fig.~\ref{Feynman}\,{\em c,\,d,\,e}, as
\begin{equation}
    \sigma^{\rm RES}  \sim
               \sum\limits_{k=T,L}f_k \otimes
               \sum\limits_{i,j}
         D_{i/\gamma^*_k} \otimes D_{j/p} \otimes
         \sigma_{ij} \ ,
\hspace{1cm}
\sigma_{ij}=c^{(1)}_{ij}\alpha_s^2+c^{(2)}_{ij}\alpha_s^3+\cdots \ ,
\label{reseq}
\end{equation}
where $i$, $j$ run over all partons in the photon and proton respectively,
$\sigma_{ij}$ is the partonic cross section, 
$c^{(1)}_{ij}\alpha_s^2$ defines the LO resolved photon cross section, 
$~c^{(1)}_{ij}\alpha_s^2+c^{(2)}_{ij}\alpha_s^3~$ the NLO resolved
photon cross section
and $f_T$, $f_L$ denote the fluxes of
transversely and longitudinally polarised virtual photons,
respectively:
\begin{eqnarray}
f_T(y,Q^2) & = & \frac{\alpha}{2\pi}
\left[\frac{2(1-y)+y^2}{y}\frac{1}{Q^2}-\frac{2m_e^2 y}{Q^4}\right],
\label{tranflux}\\
f_L(y,Q^2) & = & \frac{\alpha}{2\pi}
\left[\frac{2(1-y)}{y}\frac{1}{Q^2}\right].
\label{longflux}
\end{eqnarray}
The final dijet cross section is then given by
the sum\footnote{Care must be taken when adding the contribution 
of the LO resolved
photon diagram (Fig.~\ref{Feynman}\,{\em c}) to the NLO direct photon term
(Fig.~\ref{Feynman}\,{\em b}). To avoid double counting,
the so called photon splitting term
must be subtracted from the NLO direct photon 
contribution of the diagram in Fig.~\ref{Feynman}\,{\em b}.
%before adding the contribution of the resolved photon
%diagram of Fig.~\ref{Feynman}\,{\em c}.
%
In this paper the term ``direct contribution''
denotes the direct photon contribution before the subtraction of the
splitting term.}
of $\sigma^{\rm DIR}$ and $\sigma^{\rm RES}$.

There is an interesting connection between the direct and
resolved photon contributions.
In a large part of the phase space, the NLO direct calculations
(Fig.~\ref{Feynman}\,{\em b})
can be reasonably well approximated by
the sum of the LO direct (Fig.~\ref{Feynman}\,{\em a}) 
and LO resolved (Fig.~\ref{Feynman}\,{\em c}) photon
contributions, provided
the simplest expression, namely that given by the pure QED
splitting of the photon into a $q\overline{q}$ pair,
is used for the photon PDF~\cite{Thesis}.
In our kinematic region and for quark masses $m_q^2 \ll Q^2$,
the pure QED photon PDFs have the form
%%%
\begin{eqnarray}
D_{q_i/\gammaT}^{\mathrm{QED}}
(\xg,Q^2,E_T^2)&=&
\frac{\alpha}{2\pi}
3e_i^2
\left(\xg^2+(1-\xg)^2\right)\ln\frac{E_T^2}{\xg Q^2},
\label{QEDT}\\
D_{q_i/\gammaL}^{\mathrm{QED}}
(\xg,Q^2,E_T^2)&=&
\frac{\alpha}{2\pi}
3e_i^2
4\xg(1-\xg) \left(1-\frac{\xg Q^2}{E_T^2}\right) ,
\label{QEDL}\\
D_{g/\gamma^*_{T,L}}^{\mathrm{QED}}(\xg,Q^2,E_T^2)&=&
0 .
\label{QEDg}
\end{eqnarray}
In Eqs.~(\ref{QEDT})-(\ref{QEDg}),
$e_i$ denotes the electric charge of the quark $q_i$ 
and \xg denotes the four-momentum fraction
of the photon carried by the quark.
The full expressions for the distribution 
functions from Eqs.~(\ref{QEDT})-(\ref{QEDg}), 
containing the exact $Q^2$ dependence 
with the correct threshold
behaviour for $Q^2/m_q^2 \rightarrow 0$, can be found 
in~\cite{my}.

\subsection{The CCFM Approach}
\label{CCFM_approach}
The CCFM~\cite{CCFM} approach uses the more general concept of
an ``unintegrated'' PDF of the proton in the region of small Bjorken-$x$.
The virtualities and transverse momenta of the propagating
partons are no longer ordered, as is the case for DGLAP evolution.
Instead, an angular ordering of emissions is introduced
in order to correctly treat gluon coherence 
effects~\cite{CCFM}.
Similarly to the case of the DGLAP scheme in Eq.~(\ref{direq}), 
the cross section can be factorised into a partonic cross section
and universal parton distribution functions according 
to~\cite{Andersson:2002cf}
\begin{equation}
\sigma^{k_T \mathrm{\ FACTORISATION}}= \sum\limits_j \int \frac{dz}{z} \, d^2k_T \; 
   \hat{\sigma}_{ej}(\frac{x}{z},k^2_T)\; A_{j/p}(x,k^2_T,\muffsq)\ ,
\label{sigmaCCFM}
\end{equation}
where the partonic cross sections $\hat{\sigma}_{ej}$ have to be taken 
off-shell (i.e. dependent on the parton transverse momentum, $k_T$), 
\muff\ is the factorisation scale related to the maximum angle
allowed in the evolution,
and the unintegrated parton distributions,
$A_{j/p}(x,k^2_T,\muffsq)$, 
depend on an additional variable, the $k_T$
of parton~$j$.

The CCFM evolution scheme
provides a framework for the implementation 
of $k_T$-unordered initial state QCD cascades. 
The partons with the largest $k_T$ may 
come from any emission in the proton cascade, not necessarily 
from the hard subprocess as in the DGLAP framework. 
This can lead to events which have a similar
topology to that of the resolved photon interaction in the DGLAP
approximation~\cite{smallx},
where hard partons are accompanied by softer partons from the photon remnant.

The mean value of the proton momentum
fraction $x_p$, appearing as an argument of the unintegrated PDFs,
is $\langle x_p\rangle \simeq 0.03$ in our kinematic region.
Even though this value may not be small enough for the CCFM approach to be 
superior to that based on the standard integrated PDF and DGLAP
evolution equations, it is interesting to compare 
the CCFM predictions with the data. Recently, such comparisons 
became possible using the CASCADE Monte Carlo (MC) generator.

\subsection{Programs for Dijet Calculations}
\label{MC}
%\begin{description}
Several 
%DGLAP-based 
MC programs can be used for predictions
of dijet cross sections, as discussed below.
\begin{description}
\item{\bf HERWIG} \cite{herwig}
is a general purpose LO MC event generator,
applicable to a wide range of hard processes and
collisions, including direct and 
resolved photon interactions in the region of moderate $Q^2$.
It is based on LO cross sections in the DGLAP approach, interfaced to leading
log parton showers. 
Hadronisation is done via the decay
of colourless clusters, formed during the hard scattering and parton
shower stages. 
HERWIG is also able to model additional soft remnant-remnant
interactions  (the ``soft underlying event''), accompanying the
hard scattering process.  
The probability that a resolved photon event contains soft underlying
activity has been adjusted so that the
energy flow in and around the jets is well described.

Only transversely polarised photons are included
for resolved photon interactions in the default version of HERWIG.
To investigate the contributions of resolved
longitudinal photons, we have modified HERWIG by adding the
option of simulating the flux in Eq.~(\ref{longflux}).
Similarly the QED PDFs of the photon from 
Eqs.~(\ref{QEDT}-\ref{QEDg}) have been implemented
in HERWIG in order to study the differences between the results
obtained with the pure QED and the QCD-improved photon PDFs.
The scales \mur\ and \muf\ are set to a combination of
Mandelstam variables \cite{herwig}, which roughly corresponds to $1.1 p_T$,
where $p_T$ denotes transverse momentum of the outgoing parton
from the hard interaction.
\item{\bf RAPGAP} \cite{rapgap}
combines standard LO hard scattering matrix elements in the DGLAP approach
with parton showers and LUND
string fragmentation, using JETSET~\cite{JETSET,PYTHIA}. Only the transverse
virtual photon is considered in resolved photon processes.
Soft underlying interactions are not modelled.
\item{\bf DISENT} \cite{DISENT}
is a NLO DGLAP program for calculating dijet cross sections at
the parton level. It is based on the dipole subtraction
method~\cite{Subtr_meth} to regularise soft and collinear divergences.
The factorisation scale, \muf, was set to
$\langle E_T\rangle = 9 \, \GeV$,
since the program does not allow the user to set $\muf=E_T$
for each point in phase space.
In our kinematic region, the difference between
the results obtained with $\muf=E_T$ and
$\muf=\langle E_T\rangle$, tested
using JETVIP, is very small.
DISENT does not include resolved photon interactions.
\item{\bf JETVIP} \cite{JETVIP,JETVIP2.1}
is a NLO DGLAP parton level program which calculates both 
direct and resolved photon contributions.
It is based on the phase space slicing method.
We have performed systematic investigations of the stability of
JETVIP calculations with respect to variations of the slicing parameter 
\ycut~\cite{Thesis}.
The  direct contribution in our kinematic region is
independent of $y_c$ to within 5\% over 
the recommended range of its values $10^{-4}\le y_c\le 10^{-2}$.
The situation changes in the
case of the NLO resolved photon contribution,
for which the dependence on the \ycut parameter is significantly
larger.
The sum of NLO direct and NLO resolved JETVIP predictions
varies by 30\% in some bins for the recommended range of~\ycut.
We set $\ycut=0.003$
in all JETVIP calculations, since the predictions
are most stable around this value.

We have observed non-negligible differences
between the differential dijet cross section obtained with 
DISENT and the direct contribution from JETVIP (see 
Section~\ref{sec:datavsNLO}).
We have checked that this discrepancy is not caused by
different input parameters or kinematic cuts, as the leading
order, i.e. $O(\alpha\alpha_s)$, contributions agree perfectly.
We therefore test both programs in our analysis.
%\end{description}

\item{\bf CASCADE} \cite{CASCADE,Unintegr} is a LO MC event generator.
It uses unintegrated gluon
distribution functions of the proton, satisfying the CCFM equation,
and correspondingly produces a $k_T$ unordered initial state parton shower.
The LUND string model is used for fragmentation. 
\end{description}

%%%%\subsection{Monte Carlo event generators}
%%%\subsection{Note on Monte Carlo programs}
%%%\subsection{Monte Carlo programs and QCD Calculations}
%\subsection{Hadronisation Corrections and Specifics of MC Programs}
%\label{MC}
The parameter settings of all MC programs and NLO calculations are summarised
in Table~\ref{tabulkapar}.

\begin{table}[hbt]
\begin{center}
\begin{tabular}{lccccc}
{\bf Parameters} & {\bf HERWIG} & {\bf RAPGAP} & {\bf CASCADE} & {\bf DISENT} & {\bf JETVIP} \\
\hline \hline 
Version & 6.4 & 2.8 & 1.2 & --- & 2.1 \\
\hline 
Proton PDF & CTEQ5L~\cite{cteq} & CTEQ5L & J2003 (set 1)~\cite{Unintegr2} & CTEQ6M & CTEQ6M \\
Photon PDF & SAS1D~\cite{sas}; & SAS1D  & --- & --- & SAS1D \\
  &\cite{gammal}~for \gammaL & & & & \\
\hline 
Formula for $\alpha_s$ & one-loop & one-loop & one-loop & two-loop & two-loop \\
Active flavours & 5 & 5 & 4 & 5 & 5 \\
PRSOF & 10\% & --- & --- & --- & --- \\
\hline 
\mur & $\sim 1.1 p_T$ & $\surd (p_T^2 + m_q^2)$ & $\surd (p_T^2 + m_q^2)$ & \Etone & \Etone \\
\muf & $\sim 1.1 p_T$ & $\surd (p_T^2 + m_q^2)$ & given by & 9 GeV & \Etone \\
     &                &                          & ang. ordering&   &        \\
\hline 
Hadronisation     & Cluster model & LUND string  & LUND string  & --- & --- \\
mechanism     &               &fragmentation &fragmentation &  &  \\
\hline \hline 
\end{tabular}
\caption{Parameters of the MC programs. The variable $p_T$ denotes 
the transverse momentum of the parton with mass $m_q$ outgoing from 
the hard interaction and \Etone is the energy of the jet with the highest
transverse energy. The parameter PRSOF specifies the
fraction of resolved photon events with soft underlying
activity.} 
\label{tabulkapar}
\end{center}
\end{table}

In the NLO calculations, JETVIP and DISENT, the
massless partons entering the hard process
are taken to be exactly collinear with the beam particles.
On the other hand, the Monte Carlo generators HERWIG, RAPGAP and CASCADE
generate initial state QCD parton showers, which influence the 
four-momenta of partons entering the hard process 
by generating the appropriate transverse momentum. 
Both initial and final state parton showers also
provide more partons in the final state.
These effects tend to produce more 
low $E_T$ jets in the LO models than in the NLO calculations.

\subsection{Hadronisation Corrections}
The MC event generators have a clear advantage over the parton
level calculations in incorporating hadronisation effects.
In order to estimate the hadronisation corrections
to the NLO calculations, we use
two different Monte Carlo models, 
HERWIG and LEPTO~\cite{LEPTO},
and divide the cross sections obtained from these models for the
complete hadronic final state by the
cross sections predicted from the partonic final state
after the initial and final state QCD parton
showers.
The hadronisation corrections determined by HERWIG also include
corrections for the soft underlying event.
The average values of the corrections obtained with
HERWIG and LEPTO are applied to
the NLO calculations as bin-by-bin correction factors and
half the difference between the corrections
obtained with the two models is taken as a hadronisation
uncertainty in the NLO predictions. 
The hadronisation effects usually do not change the
NLO predictions by more than 5\%, with the exception of
the cross section differential in $\xg$,
for which the corrections change the cross section by
up to~15\%.

%%%%%%%%%%%%%%%%%%%%%%%%%%%%%%%%%%%%%%%%%%%%%%%%%%%%%%%
%
%
\section{Detector Description}
\label{detector}
A detailed description of the H1 detector can be found
elsewhere~\cite{h1det} and only the components relevant for
this analysis are described here.

The H1 central tracking system is mounted concentrically around 
the beam-line and covers polar 
angles\footnote{The $z$ axis of the right-handed coordinate system
used by H1 is defined to lie along the direction of the proton
beam with the origin at the nominal $ep$ interaction vertex.}
$20^\circ < \theta < 160^\circ$.
The transverse momenta and charges of charged
particles are measured by two coaxial cylindrical
drift chambers~\cite{cjc}. 
Two drift chambers which provide accurate measurements of the $z$
coordinate of charged tracks and two multi-wire proportional chambers
which trigger on these tracks
are placed on either side of the inner main drift chamber. 

The tracking system is surrounded  by
a finely segmented Liquid Argon Calorimeter~\cite{larcalo},
which covers the range of polar angles 
$4^\circ < \theta < 154^\circ$ and the full range in azimuth. 
It consists of an electromagnetic section with lead absorbers,
20-30 radiation lengths in depth, and a hadronic section with
steel absorbers. The total depth of the calorimeter ranges 
from 4.5 to 8 hadronic interaction lengths. The energy
resolution obtained from test beam measurements~\cite{testcalo}
is 
$\sigma(E)/E \approx 0.11/\sqrt{E}$ for electrons and 
$\sigma(E)/E \approx 0.5/\sqrt{E}$ for
pions, with $E$ in GeV.
The absolute energy scale 
for hadrons is known for this analysis to a precision of 4\%.
%electrons and 
A uniform axial magnetic field of 1.15 T
is provided by a superconducting coil, which surrounds the 
calorimeter. 

The polar angle region $153^\circ < \theta < 177.8^\circ$ is
covered by the SPACAL~\cite{spacal}, a lead/scin\-til\-lat\-ing
fibre calorimeter with electromagnetic
and hadronic sections. The energy resolution of
the electromagnetic section is determined to be
$0.07/\sqrt{E}\oplus 0.01$ ($E$ in GeV)~\cite{Nicholls:1995di}. 
Both calorimeter sections have a time resolution better
than 1~ns.
The SPACAL is used both to trigger on the scattered 
electron\footnote{In the following, the notation ``electrons''
stands for both positrons and electrons.}
and to measure its energy.
In front of the SPACAL,
an eight layer drift chamber, BDC~\cite{bdc}, covers the polar
angle region $151^\circ < \theta < 177.5^\circ$.
It is used to suppress background from neutral particles faking
the scattered electron and, together with the vertex obtained
from the central drift chambers, to measure the scattered electron
polar angle~$\theta$.

%%%%%%%%%%%%%%%%%%%%%%%%%%%%%%%%%%%%%%%%%%%%%%%%%%%%%%%
%
%
\section{Data Samples and Event Selection}
\label{cuts}
The present analysis is based on data taken in the years 1999 and 2000,
when electrons with an energy of 27.55~GeV collided with protons with an
energy of 920~GeV.
The data correspond to an integrated luminosity of \lumi.
83\% of the data sample corresponds to $e^+p$ collisions, the remainder
to $e^-p$ interactions.

The kinematic region covered by the analysis is defined by cuts
on the photon virtuality, $Q^2$, the inelasticity, $y$, and by cuts 
on the hadronic final state. We require
$$2< Q^2 < 80~{\GeV}^2,$$
\begin{equation*}
0.1 < y < 0.85 .
\end{equation*}
The variables \qq and $y$ were determined using the scattered electron 
energy and polar angle~\cite{Tanja}.

The final state has to contain at least two jets.
Jets are found using the longitudinally invariant $k_t$ jet
algorithm~\cite{INVKT} applied to hadronic final state
``combined objects'', boosted into the photon-proton 
centre-of-mass frame. 
The combined objects are constructed from tracks in the central 
track chambers and clusters in the SPACAL and LAr calorimeters
in a procedure that avoids double counting~\cite{FSCOMB}.
The jet transverse energies, \Etmy\!, 
and pseudorapidities, \etamy\!, are calculated relative to 
the $\gamma^* p$ collision axis in the $\gamma^* p$ centre-of-mass 
frame\footnote{The pseudorapidity is defined by
$\etamy \equiv-\ln(\tan\theta^*/2)$, 
where $\theta^*$ is the polar angle of the jet 
axis with respect to the $\gamma^* p$ collision axis. 
Negative values of \etamy correspond to the photon fragmentation region.
The pseudorapidity in the photon-proton centre-of-mass frame is shifted
on average by -2.3 units with respect to the pseudorapidity in 
the laboratory frame.}.
The jets are ordered according to their
transverse energy, with jet 1 being the highest \Etmy jet.

The two jets with the highest transverse energies (leading jets) 
are required to have
$$\Etone > 7~\GeV,~~~ \Ettwo > 5~\GeV,$$
\begin{equation*}
-2.5 < \etaone < 0,~~~ -2.5 < \etatwo < 0.
\end{equation*}
The asymmetric \Etmy cuts 
avoid regions of instability in the
NLO calculations~\cite{Etcut,Chyla:2003in}.

The reconstructed event vertex has to be within $\pm 35$ cm
of the nominal interaction point, which substantially reduces
contributions from beam induced background.
To remove background from photoproduction processes,
a cut $45<\sum_{i}(E_i-p_{z,i})<75\;\GeV$ is applied, where the sum runs over
all particles in the final state including the scattered electron. 
In total 105\,658 events satisfied the selection criteria.

It is convenient to describe dijet events by means of
the variable $\xgmy$, defined as
\begin{equation}
\label{rovnice3}
\xgmy = \frac{\sum\limits_{j=1,2}(E^*_{j}-p^*_{z,j})}
{\sum\limits_{\rm hadrons}(E^*-p^*_{z})}\ ,
     \end{equation}
where the sum in the numerator runs over the 
two leading jets and the sum in the 
denominator includes the full hadronic final state.
Neglecting the masses of the partons and beam particles,
the variable \xgmy 
represents a hadron level estimate of the fraction of the 
photon four-momentum carried by the parton involved in the 
hard scattering.

%%%%%%%%%%%%%%%%%%%%%%%%%%%%%%%%%%%%%%%%%%%%%%%%%%%%%%%
%
%
\section{Analysis Procedure}
\label{systematics}
The data were corrected for initial
and final state QED radiation effects using samples of RAPGAP
events for direct photon interactions with
and without QED radiation, processed through the full
detector simulation and fulfilling all the cuts described in the
previous section. 
The effects of trigger inefficiencies,
limited detector acceptance and resolution 
were corrected for using 
an iterative Bayesian unfolding technique~\cite{unfold},
which was applied
to events generated by HERWIG and RAPGAP.
For this purpose, 4 million events from each generator were passed 
through the full simulation of the H1 detector and the same chain of 
reconstruction and analysis procedure as for the data.

The binning for the final results was chosen such that the bin width 
is always larger than the resolution of the given quantity.
The iterative Bayesian procedure converged in all bins of the 
measured quantities~\cite{Thesis}.  
After unfolding, the correlations between neighbouring bins
in the unfolded distributions were always less than $60\%$.
The correction factors from the unfolding procedure 
were cross-checked with a bin-by-bin 
correction method, performed using the same simulated MC samples,
and agreement was found to within 5\%.

In all of the distributions studied, 
the presented cross sections are taken as averages of the cross 
sections obtained when correcting for detector effects using HERWIG and RAPGAP,
since the description from the two models of 
the uncorrected distributions are of similar quality.

The background in the event sample from photoproduction events, in which 
a hadron in the SPACAL is misidentified as the electron candidate,
was estimated using PYTHIA~\cite{PYTHIA} and PHOJET~\cite{PHOJET} 
MC samples of photoproduction events.  
This background is negligible for most of the bins and reaches 4\%
at the highest $y$.

The systematic errors are added in quadrature.
They are listed below in order of their size:
\begin{itemize}
    \item[--] {\it Model Dependence.}
       The systematic error from the model dependence of the acceptance
       corrections is taken as half of the difference between the 
       results when unfolded with RAPGAP and with HERWIG.
       This leads to an error of 5--10\% on average, reaching 
       20\% in the most extreme case.
    \item[--] {\it Energy Calibration of the Calorimeters.}
       Varying the overall hadronic energy scale of the LAr calorimeter 
       by 4\%, the hadronic energy scale of the SPACAL by 7\% 
       and the electromagnetic energy scale of the SPACAL 
       by 1\% leads to systematic shifts
       of the results by typically 10\%, 2\% and 4\%, respectively.
    \item[--] {\it Scattered Electron Angle.}
       The polar angle of the scattered electron
       is measured with a precision of 1~mrad, which leads to a 3\% (1\%)
       systematic uncertainty in the lowest (highest) $Q^2$ region. 
    \item[--] {\it Trigger Efficiency.} The uncertainty in the
       trigger efficiency leads to a 3\% uncertainty in the measurement.
    \item[--] {\it QED Radiative Corrections.}
       2\% is taken as the systematic error in all bins~\cite{Thesis}.
    \item[--] {\it Stability of the Bayesian Unfolding Procedure.}
       By varying the number of iterations used in the unfolding procedure, 
       the uncertainty due to the unfolding instability is estimated 
       to be typically less than 2\% and at most 5\%.
    \item[--] {\it Photoproduction Background.}
       The photoproduction background is subtracted statistically 
       and half of the subtracted background is taken as the systematic 
       uncertainty.
    \item[--] {\it Precision of the Luminosity Measurement.}
       The normalisation uncertainty due to the
       luminosity measurement is 1.5\%.
\end{itemize}

%%%%%%%%%%%%%%%%%%%%%%%%%%%%%%%%%%%%%%%%%%%%%%%%%%%%%%%
%
%
\section{Results}
\label{results}
Differential cross sections for the kinematic region
defined in Section~\ref{cuts} are discussed in the
following sections and
presented in Tables~\ref{tabqe3x}-\ref{tabqx4y}.

\subsection{Comparison with NLO Calculations}
\label{sec:datavsNLO}
\begin{figure}\centering
\epsfig{file=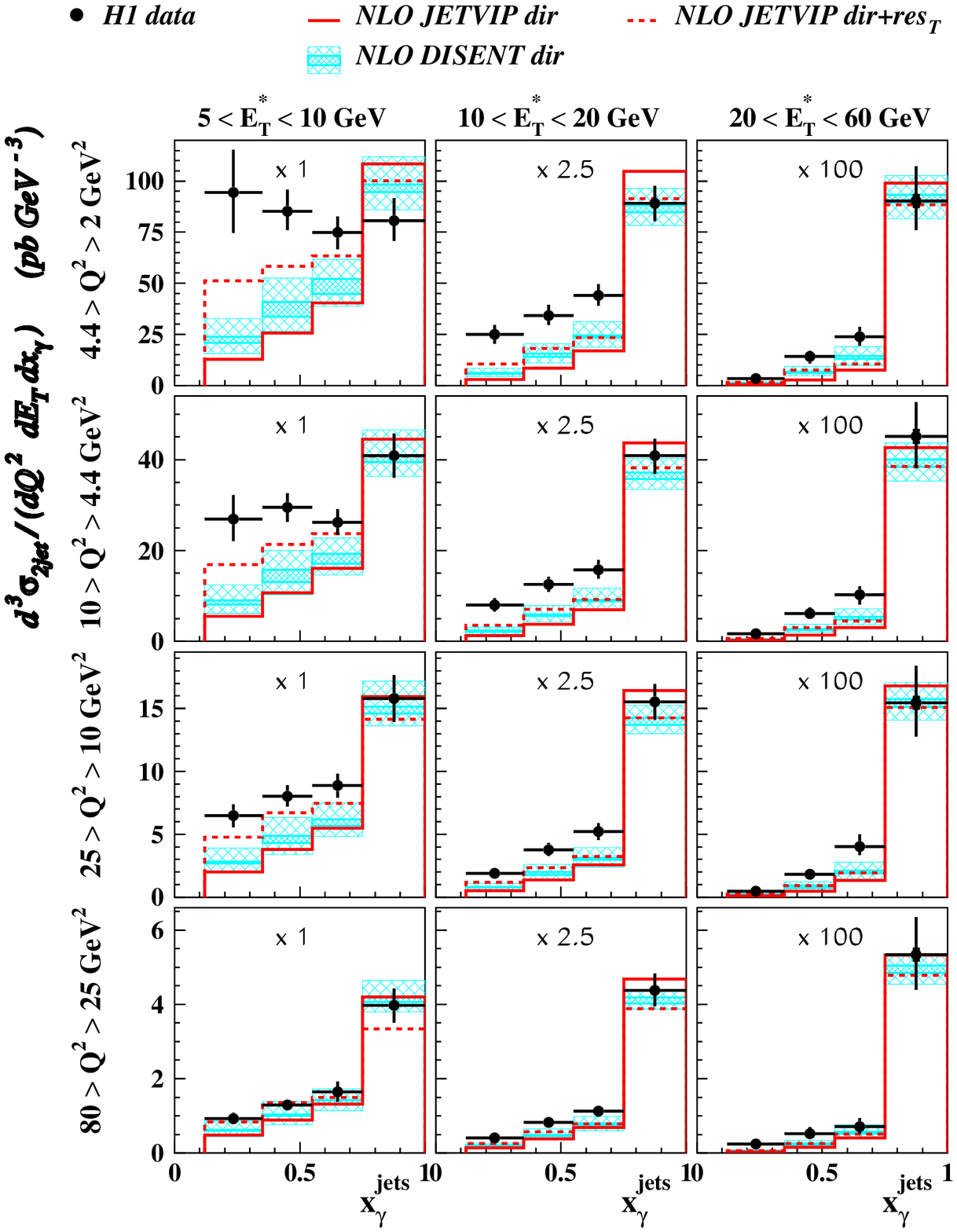,height=19cm,%
bbllx=5pt,bblly=15pt,bburx=520pt,bbury=673pt,clip=}
\caption{Triple differential dijet cross section,
      \dsqex\!, with asymmetric \Etmy cuts (see text).
      The inner error bars on the data points show the statistical error, 
      the outer error bars show the quadratic sum of systematic
      and statistical errors.
      Also shown are NLO direct photon calculations using DISENT 
      (hatched area) and JETVIP (full line), 
      as well as the sum of NLO direct and NLO resolved photon contributions
      of JETVIP (dashed line). All calculations are corrected for 
      hadronisation effects.
      The inner hatched area illustrates
      the uncertainty due to the hadronisation corrections,
      the outer hatched area shows the quadratic sum of the
      errors from hadronisation and the scale uncertainty (shown
      only for DISENT).
      The scale factors applied to the cross sections are given.}
\label{qe3x.dis}
\end{figure}

The triple differential dijet
cross section is presented as a function of \xgmy in different 
bins of \qq and \Etmy
in Fig.~\ref{qe3x.dis}. The variable \Etmy 
denotes the transverse energies of the jets with the highest and second highest
\Etmy measured in the photon-proton centre-of-mass frame, 
so that each event contributes twice to the
distributions, not necessarily in the same bin. The data are
compared with the NLO direct photon 
calculations\footnote{The resolved photon prediction of JETVIP,
also shown in Fig.~\ref{qe3x.dis},
is discussed in Section~\ref{sec:datavsRES}.}
performed with DISENT and JETVIP.
The uncertainties from variations of the 
factorisation and renormalisation scales in the interval
$\mu/2$ to $2\mu$,
as well as from hadronisation corrections, are illustrated.
The scale uncertainties are typically around 20\%. 
Those from hadronisation are at the 7\% level.
We have also investigated the uncertainties due to variations of the
proton PDF, using the prescription of~\cite{Pumplin:2002vw}.
The typical uncertainties are below 4\% (not shown).
Figure~\ref{qe3x.dis} demonstrates that the NLO direct photon 
calculations describe the data in the region of high \xgmy\!, where
direct photon interactions dominate.
For $\xgmy < 0.75$, the description
is nowhere perfect, indicating the need for orders beyond NLO. 
The description of the data for $\xgmy < 0.75$ gets
worse as $Q^2$ and \Etmy decrease.
The discrepancy is particularly pronounced at small
\mbox{\xgmy\!,} low $Q^2$ and low \Etmy\!, 
where the data lie significantly above the
theoretical predictions, even taking into account the sizable
scale uncertainty. 
The relative decrease of the cross section at low \xgmy 
as \Etmy increases is of kinematic origin, due to the restrictions 
in the available phase space.
Note that for $\xgmy < 0.75$, the JETVIP
results are systematically lower than those of DISENT, 
whereas for $\xgmy >0.75$ the opposite effect is observed.
The discrepancy between
DISENT and JETVIP is observed only for 
multi-differential distributions which include a jet variable.
It gets substantially smaller for the
inclusive dijet cross section \dsqy~\cite{Thesis} (not shown)
and agrees within 2\% for the total dijet cross section
in our kinematic region.
A similar level of agreement between JETVIP and DISENT 
was reported in~\cite{Duprel:1999wz} for the total dijet cross section.

The data were also analysed in terms of jet pseudorapidities.
Figure~\ref{qy1r5.dis} presents the dijet cross section as
a function of \etamy in different bins of \qq and $y$, 
where \etamy denotes the
pseudorapidities of the jets with the highest and second highest \Etmy
in the photon-proton centre-of-mass system, such 
that each event enters the distributions twice.
The excess of the data over the theory at low $Q^2$ and low $\xgmy$
observed in Fig.~\ref{qe3x.dis}
is reflected in Fig.~\ref{qy1r5.dis}
in a similar excess at low $Q^2$ and high $y$,
which is especially pronounced
in the forward region of the laboratory frame ($\etamy \sim 0$). 

\begin{figure}\centering
\epsfig{file=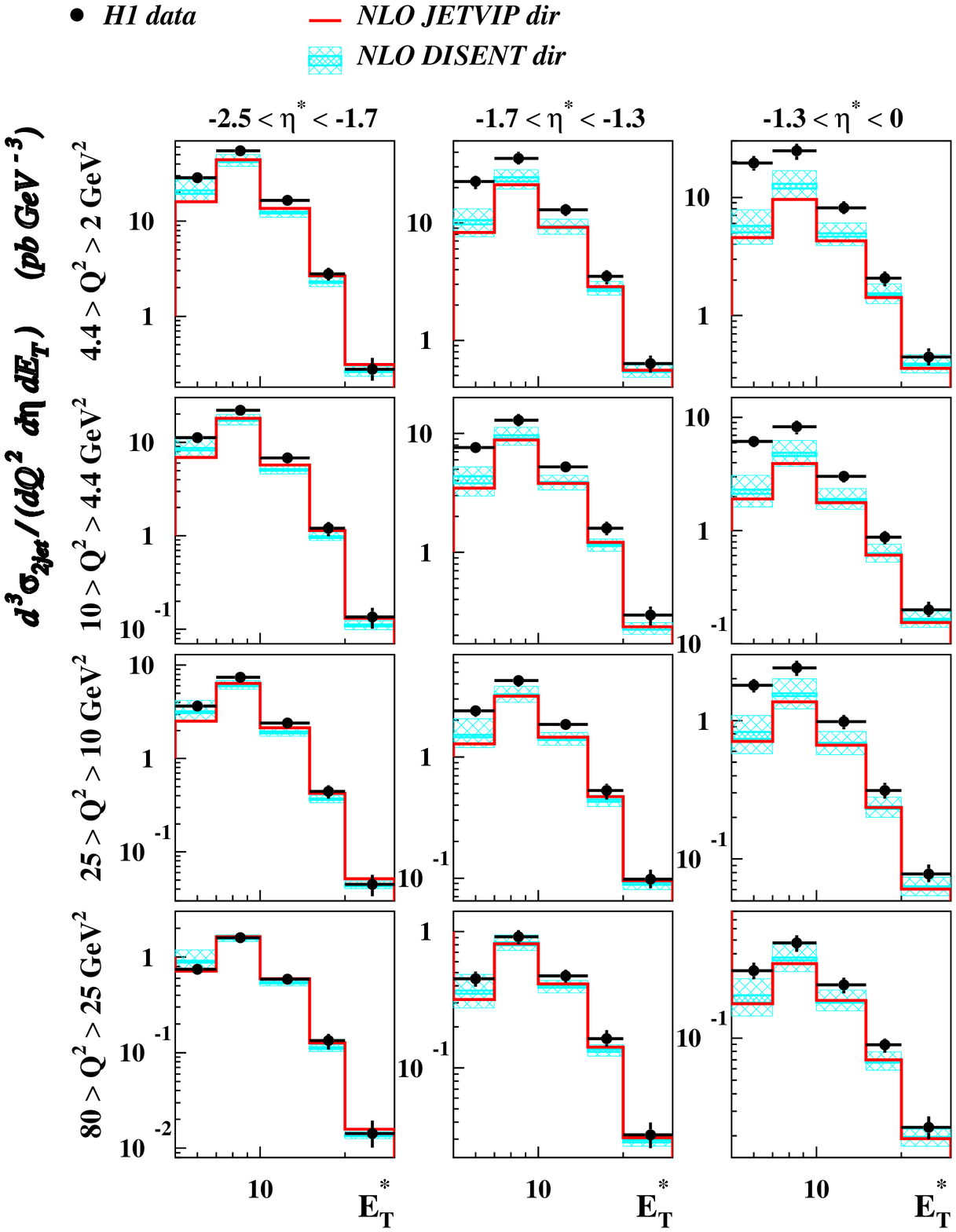,height=19cm,%
bbllx=5pt,bblly=15pt,bburx=520pt,bbury=673pt,clip=}
\caption{Triple differential dijet cross section, \dsqre\!.
        Negative values of \etamy correspond to the photon
        fragmentation region.
        See the caption of Fig.~\ref{qe3x.dis} for further details.}
\label{qre1.dis}
\end{figure}

Figure~\ref{qre1.dis} shows the triple differential dijet cross
section as a function of \Etmy in different bins of \qq
and \etamy\!, each event entering the distributions twice. 
The predictions of the NLO direct 
calculations agree well with the data at large $Q^2$
or at large \Etmy for all \etamy\!. 
On the other hand, the predictions clearly fail to
describe the data in the forward region
at low $Q^2$ and low \Etmy.
The low \Etmy region is better described as \etamy is
reduced or $Q^2$ is increased.
A similar discrepancy between the data and
the NLO prediction has 
recently been reported for inclusive
jet cross sections in a similar kinematic region~\cite{Forward}.
The measurement in~\cite{Forward}
indicated that the region where the NLO calculations fail to describe 
the data
corresponds to the region where the ratio of NLO to LO predictions
is largest.  
The same is true for the dijet cross sections.
The NLO corrections are smallest in the backward region
at the largest $Q^2$ 
and \Etmy\!, where the ratio is approximately 1.1.
The data are well
described by the NLO direct calculations in this kinematic 
region, as can be seen in Fig.~\ref{qre1.dis}.  
On the other hand, the ratio of NLO to LO predictions for the 
forward region at small $Q^2$ and \Etmy becomes as large as 9 
and the data there are not reproduced
by the NLO calculations.
Corrections beyond NLO are therefore expected to improve
the description in this region.

The above comparisons show that
in the region of low $Q^2$, high $y$, forward \etamy and low \Etmy
the data lie significantly above NLO QCD calculations
for direct photons.
This excess cannot be accommodated within standard theoretical 
uncertainties from scale variations and hadronisation 
corrections.

\subsection{Resolved Virtual Photons}
\label{sec:datavsRES}
The pattern of the observed discrepancy between the data and
the NLO calculations in \mbox{Fig.~\ref{qe3x.dis}-\ref{qre1.dis}}
suggests an explanation in terms of the interactions of
resolved virtual photons, understood as an approximation 
to contributions beyond NLO.
Of the NLO parton level calculations, only JETVIP includes 
a resolved virtual photon 
contribution\footnote{Only the contribution of transversely polarised
resolved photons is implemented in JETVIP.}.
Unfortunately, 
the dependence on the slicing parameter \ycut of the NLO 
JETVIP calculations of the resolved \gammaT contribution 
(i.e. up to order $\alpha\alpha_s^3$) is much larger
than for the direct component~\cite{Thesis}, and the resulting
calculations are therefore less reliable.
In the absence of other
calculations of this kind, the data are compared
with the results of the full JETVIP
calculations in Fig.~\ref{qe3x.dis}
using $\ycut=0.003$ (see Section~\ref{MC}),
in order to see the qualitative effects of resolved
photon interactions at NLO.

The inclusion of a resolved \gammaT contribution 
brings the NLO calculations closer to the data,
though there is still a discrepancy between the data 
and calculations at low to moderate \xgmy and low $Q^2$.
The dominant part of the difference 
between the full NLO JETVIP results and the direct component comes
from the ${\cal O}(\alpha \alpha_s^3)$ term
in the resolved \gammaT contribution (Fig.~\ref{Feynman}\,{\em{e}}).
Including only the leading 
resolved \gammaT terms (Fig.~\ref{Feynman}\,{\em{c,d}}) has
only a small effect~\cite{Thesis}.
%although it goes qualitatively in the right direction.

\subsection{Comparison with DGLAP Monte Carlo Models}
\label{sec:datavsDGLAP}
\begin{figure}\centering
\epsfig{file=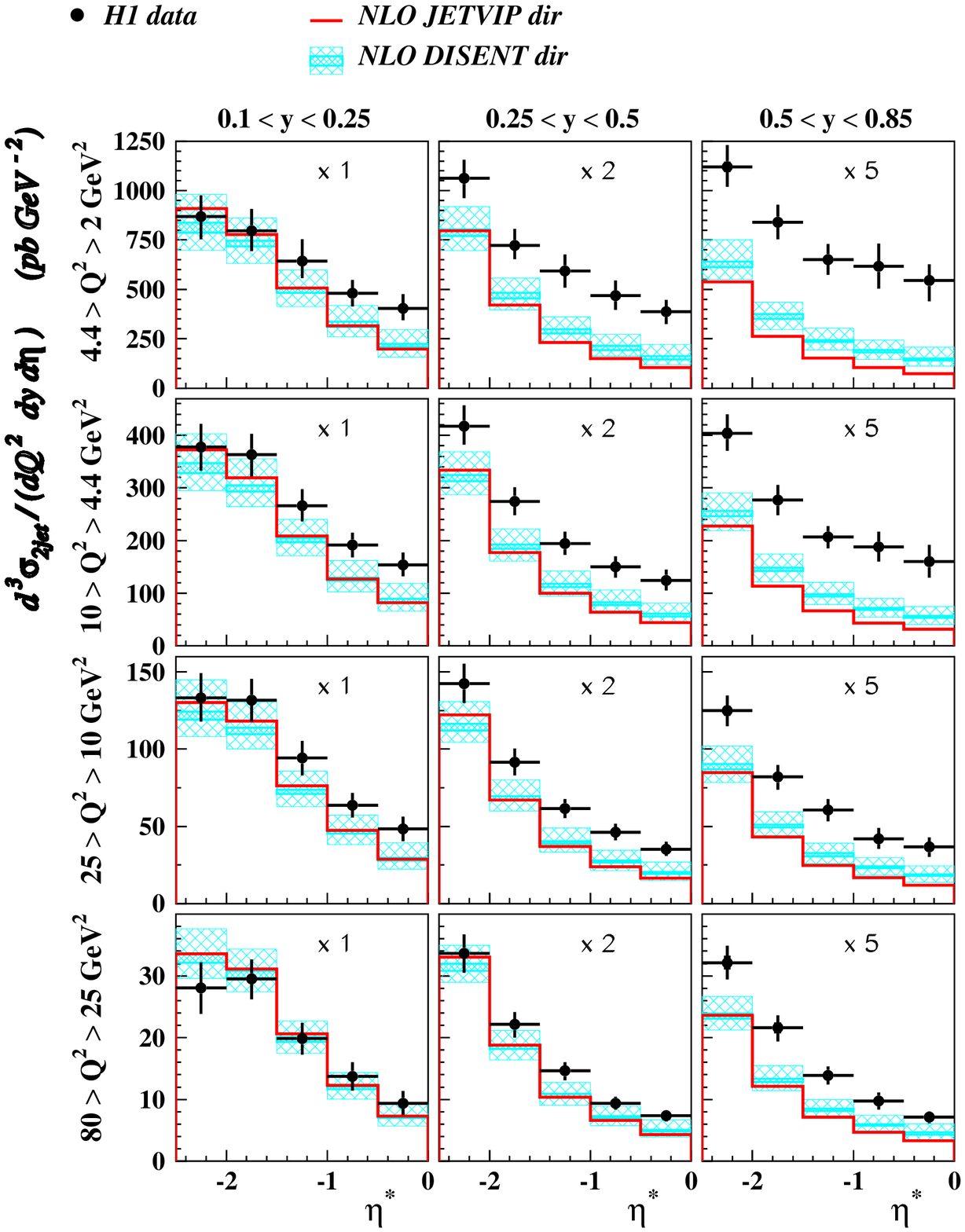,height=19cm,%
bbllx=5pt,bblly=15pt,bburx=520pt,bbury=673pt,clip=}
\caption{Triple differential dijet cross section, \dsqyr\!.
        Negative values of \etamy correspond to the photon
        fragmentation region.
        See the caption of Fig.~\ref{qe3x.dis} for further details.}
\label{qy1r5.dis}
\end{figure}

\begin{figure}\centering
\epsfig{file=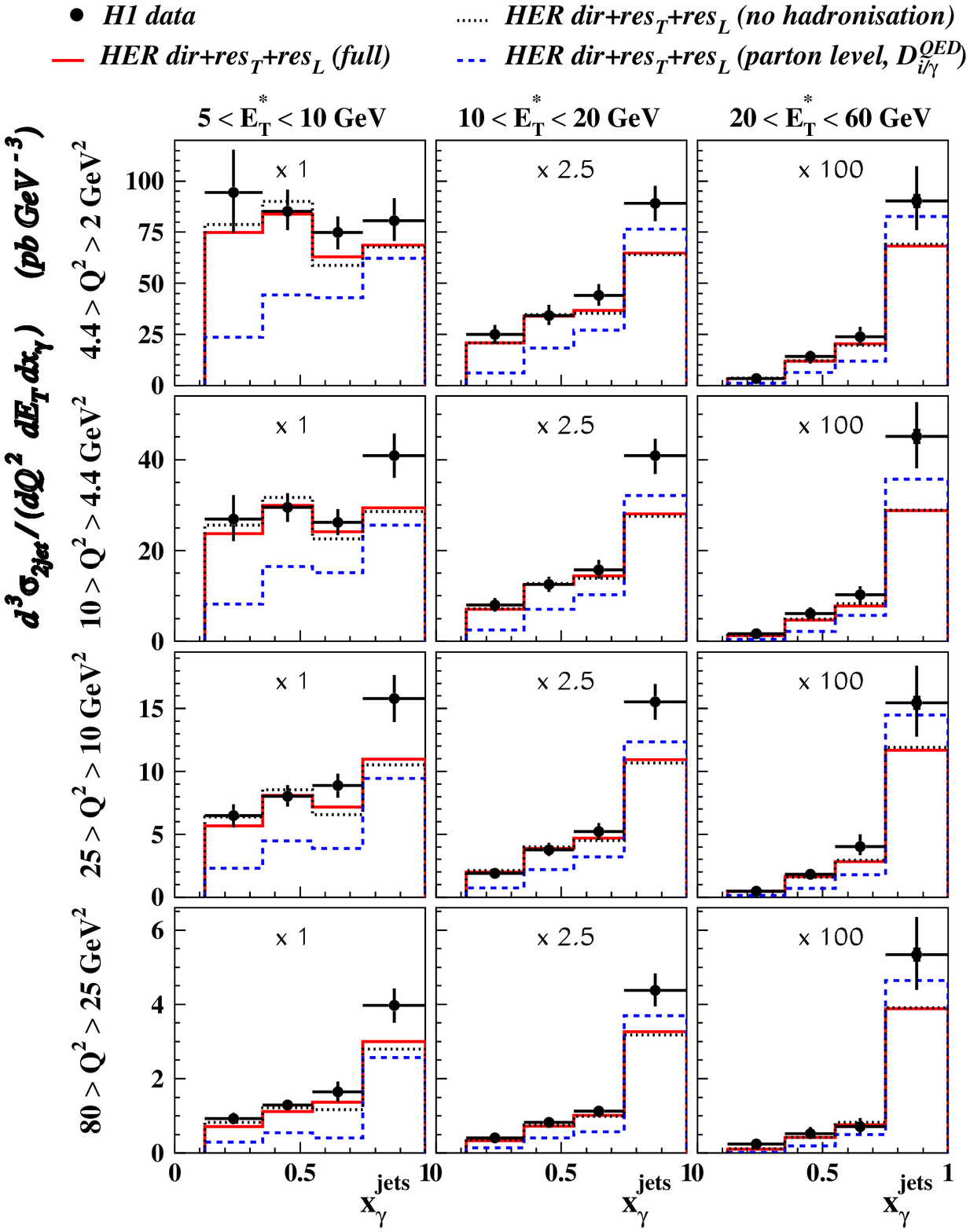,height=19cm,%
bbllx=5pt,bblly=15pt,bburx=520pt,bbury=673pt,clip=}
\caption{The triple differential dijet cross section, \dsqex\!,
      from H1 data, compared with the predictions of
      the full HERWIG simulation as defined in Section~\ref{MC} (full line),
      HERWIG without hadronisation or soft underlying
      event effects (dotted line)
      and HERWIG at the parton level 
      (dashed line). In the latter case, 
%no hadronisation, soft underlying activity or QCD parton showers 
%      have been simulated, and 
the QED PDFs of \gammaT and \gammaL 
      (see Eqs.~(\ref{QEDT}) and~(\ref{QEDg})) are used 
      in the resolved photon contributions.}
\label{qe3x.ppp}
\end{figure}

Unlike the parton level calculations, DISENT and JETVIP, discussed in 
the context of Fig.~\ref{qe3x.dis}-\ref{qre1.dis},
LO MC models take initial and final state QCD parton showers into account.
Their importance, together with the QCD improvements of \PDFgamma 
(see Eq.~(\ref{reseq})) 
and the simulation of soft underlying interactions and hadronisation,
is demonstrated in Fig.~\ref{qe3x.ppp}.  
The full HERWIG simulation, as described in Section~\ref{MC}, 
is compared with the HERWIG prediction without hadronisation or
soft underlying event effects, 
and with the HERWIG calculation at the parton level
without parton showers and with only the QED PDFs
of virtual photons.
%\footnote{Note that this parton level HERWIG prediction
%is very close to the NLO direct photon calculation, as stated in 
%Section~\ref{DGLAP_approach}.}.
The cross sections predicted
by the full HERWIG simulation are in good agreement 
with the data in the low \xgmy region.
The highest \xgmy region is not described so well.
The largest difference between the cross section predicted by the full
HERWIG simulation and that obtained from the parton level calculation
comes from the
initial and final state QCD parton showers, which effectively
introduce an intrinsic $k_T$ of the partons in the incoming proton. 
These effects increase the total dijet cross section
in our kinematic region typically by 30\% and by as much as 100\%
at low $Q^2$, low \Etmy and low $\xgmy$.
Another 10\% increase of the total dijet cross section arises 
from the change from QED to QCD-improved $\PDFgamma$.
Soft underlying events increase the total dijet 
cross section by 4\%. Their influence is largest in the
region of low $Q^2$, low \Etmy and low \mbox{\xgmy\!,} where
the cross section is increased by 10\%.
A similar effect is observed when using the multiple interaction
model implemented in PYTHIA.

Previous analyses of jet production in low $Q^2$ $ep$ collisions 
have compared with 
resolved virtual photon
models that neglect longitudinally
polarised photons.
Figure~\ref{qe3x} shows predictions of the direct and \gammaT and \gammaL 
resolved photon components separately.
At high \qq 
the HERWIG direct photon prediction alone reasonably describes the
shape of the \xgmy distribution of the data, while at low $Q^2$
the resolved photon contributions are clearly needed.
The contribution of longitudinally polarised resolved photons
improves the agreement with the data.
Not only do they increase the magnitude of the HERWIG predictions 
such that they become closer to data,
but they also correctly reproduce the
\qq and \Etmy dependence. For a given interval of \Etmy\!, the ratio of
\gammaL to \gammaT contributions increases with $Q^2$,
whereas keeping \qq fixed it decreases with increasing \Etmy.
This behaviour is expected from Eqs.~(\ref{QEDT}) and~(\ref{QEDL}).
Enhancing the PDF of
\gammaT in the resolved photon contribution by a constant factor does
not lead to a comparably successful description of the data.

As a result of the different 
$y$ dependences of the photon fluxes in
Eqs.~(\ref{tranflux}) and~(\ref{longflux}),
the dijet cross section as a function of $y$
is different for longitudinal and transverse photons.
Figure~\ref{qx4y} shows the event cross section 
as a function of $y$ in different bins of \qq and $\xgmy$.  
In contrast to all previous dijet cross sections, each event
contributes only once to the event cross section \mbox{\dsqxy}.
The ratio of longitudinally to transversely polarised photons
decreases with increasing $y$, as expected from 
the fluxes (see Eqs.~(\ref{tranflux}) and (\ref{longflux})).
The addition of the resolved longitudinal photon contribution 
brings the HERWIG 
predictions\footnote{The low HERWIG prediction for all bins 
with $\xgmy > 0.75$ and at the lowest $y$ for $\xgmy < 0.75$ 
is partially due to a cut-off procedure in HERWIG, which suppresses 
the PDF of the virtual photon at large $\xgmy$. 
The resolved \gammaT contribution of RAPGAP (not shown) leads 
to a rise with decreasing $y$ that is similar to that in the data 
for the low \xgmy range, though RAPGAP also lies below the data 
in the large \xgmy range~\cite{Thesis}.}
closer to the data.
The small contribution of \gammaL compared to \gammaT
at large \xgmy is a consequence of the 
different \xg dependences of \PDFgammaT and \PDFgammaL
(see Eqs.~(\ref{QEDT}) and~(\ref{QEDL})).

Figure~\ref{qy1r5} compares the measured dijet cross section
as a function of \etamy in different bins of \qq\! and $y$,
presented already in Fig.~\ref{qy1r5.dis}, with the HERWIG prediction.
The data are well reproduced by the complete LO MC model
in shape. However, the absolute normalisation is not satisfactory,
especially at low $y$.
In agreement with the conclusion of Figs.~\ref{qe3x} and~\ref{qx4y},
the importance of the resolved photon contributions increases in the
forward jet region ($\etamy \sim 0$), for low \qq and at high $y$.

\subsection{Comparison with CCFM Monte Carlo Model}
\label{sec:datavsCCFM}
\begin{figure}\centering
\epsfig{file=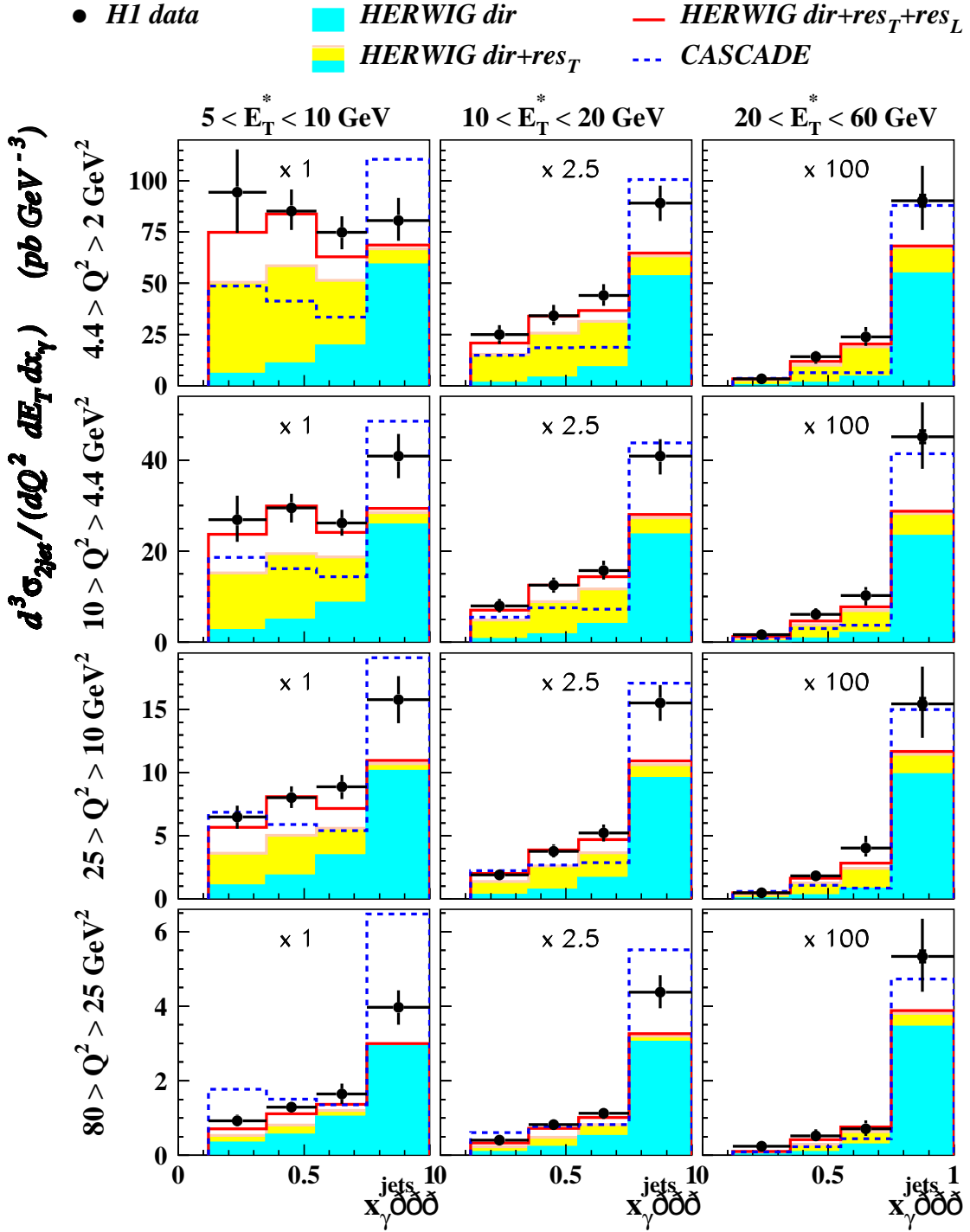,height=19cm,%
bbllx=5pt,bblly=15pt,bburx=520pt,bbury=673pt,clip=}
\caption{The triple differential dijet cross section, \dsqex\!,
      from H1 data
      compared with the predictions of HERWIG and CASCADE.
%      (dashed line).
      The dark-filled histograms show the direct HERWIG contribution.
      The light-filled histograms the resolved \gammaT HERWIG 
      prediction and the full line is the sum of all direct, \gammaT
      and \gammaL resolved HERWIG contributions.}
\label{qe3x}
\end{figure}

\begin{figure}\centering
\epsfig{file=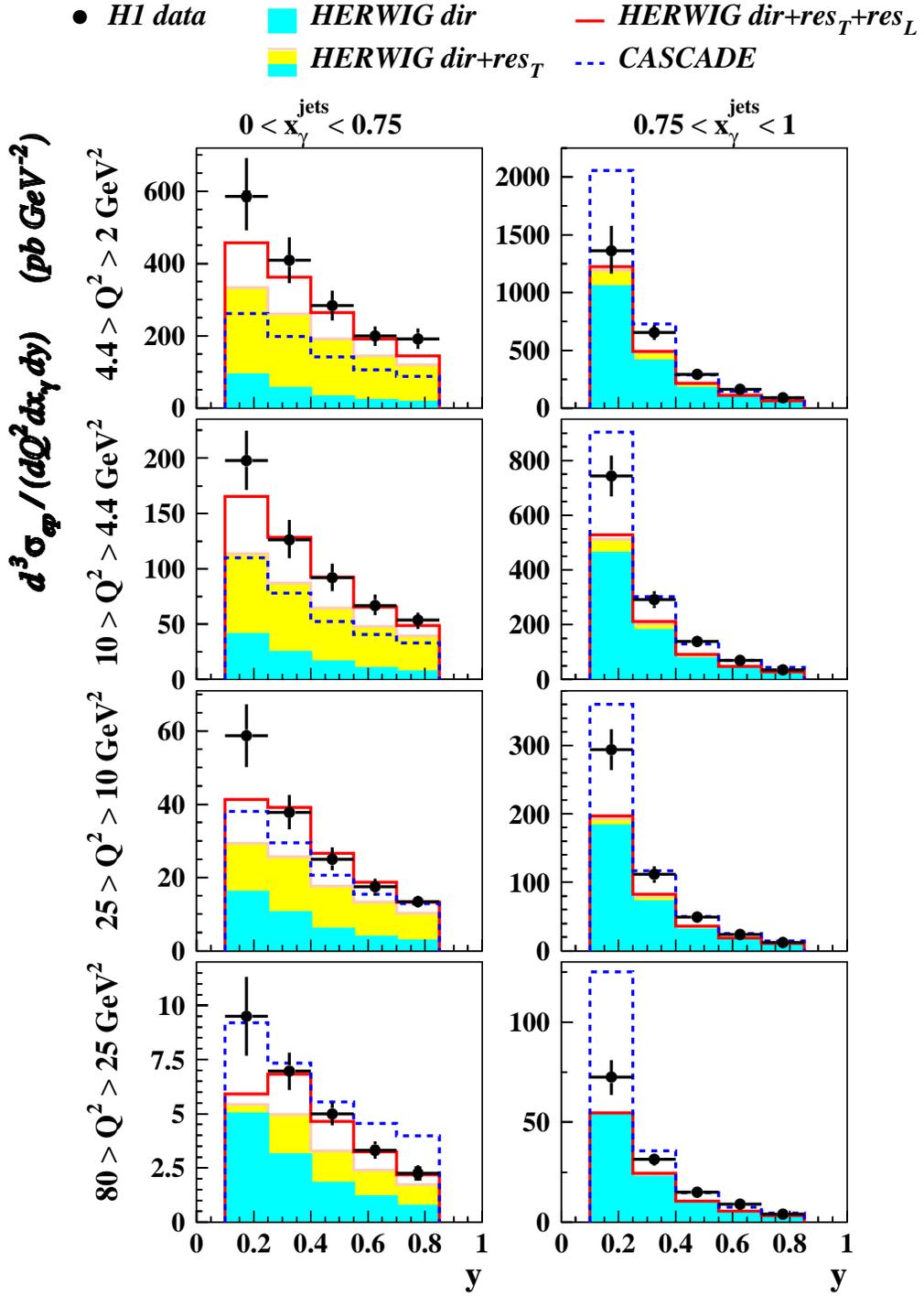,height=19cm,%
bbllx=2pt,bblly=15pt,bburx=465pt,bbury=673pt,clip=}
\caption{Triple differential event cross section, \dsqxy\!.
         See the caption of Fig.~\ref{qe3x} for further details.}
\label{qx4y}
\end{figure}

\begin{figure}\centering
\epsfig{file=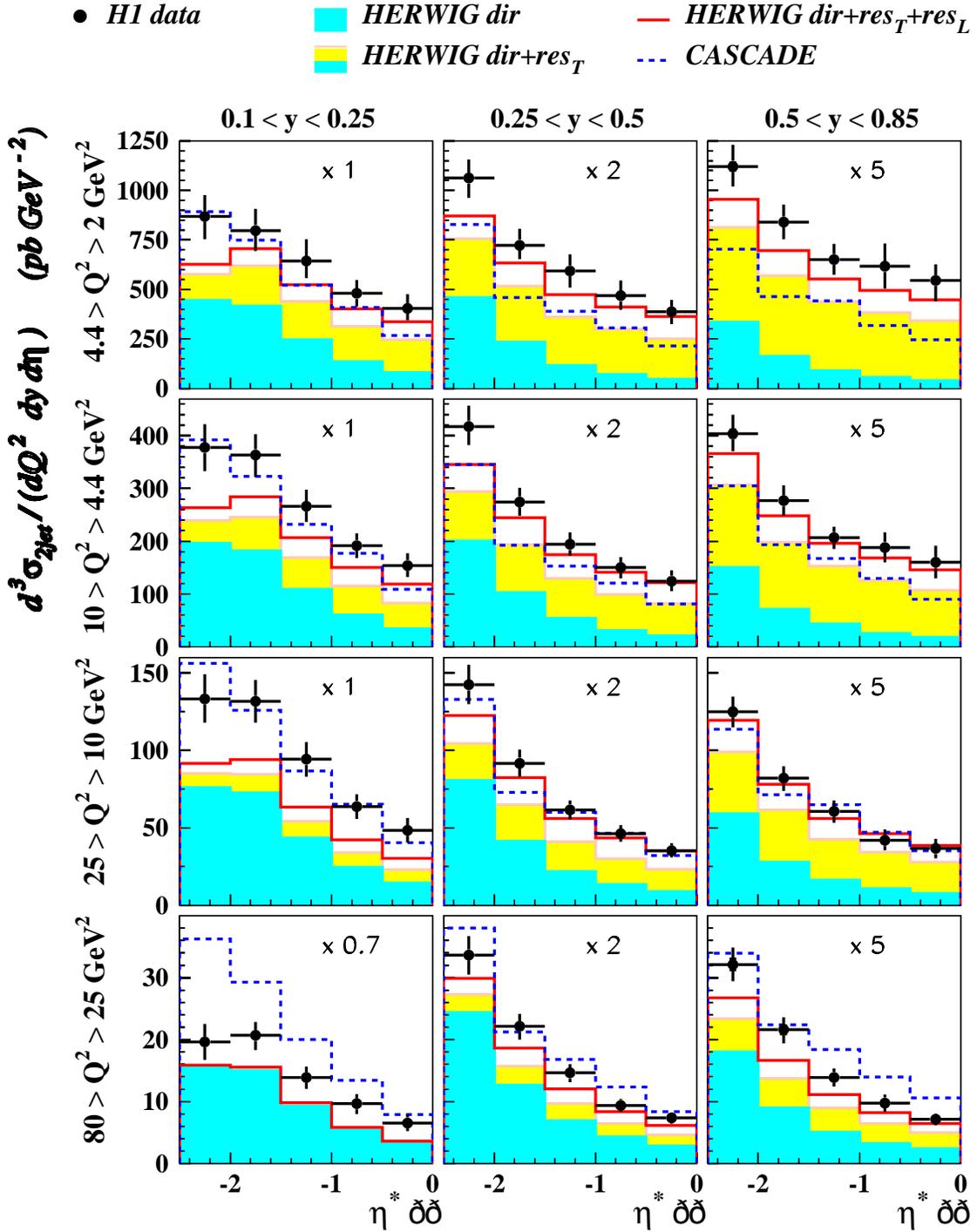,height=19cm,%
bbllx=5pt,bblly=15pt,bburx=520pt,bbury=673pt,clip=}
\caption{Triple differential dijet cross section, \dsqyr\!.
         See the caption of Fig.~\ref{qe3x} for further details.}
\label{qy1r5}
\end{figure}

In Figs.~\ref{qe3x}-\ref{qy1r5}, the data are also compared with
the predictions of the CASCADE MC, employing the unintegrated PDF 
(set~1 of~\cite{Unintegr2})
with the CCFM evolution equations. 

The CASCADE prediction
describes the main qualitative trends in the data,
except the $Q^2$ dependence in the lowest 
\Etmy bin (Fig.~\ref{qe3x})
or at low \xgmy (Fig.~\ref{qx4y}).
CASCADE also overestimates the data in the lowest 
$y$ bin at high \xgmy (Fig.~\ref{qx4y}).
On the other hand, CASCADE predicts a significant dijet 
cross section at low \xgmy (Fig.~\ref{qe3x}), much higher 
and closer to the data than the LO and NLO DGLAP predictions without 
the resolved photon interactions.
Also, except for the highest $Q^2$ bin, 
dijet production in the forward region 
is reproduced better by CASCADE (Fig.~\ref{qy1r5}) than by
NLO direct photon calculations (Fig.~\ref{qy1r5.dis}).

Large sensitivity of the CASCADE predictions to the choice of unintegrated 
proton PDF is observed~\cite{Thesis}. The results shown here are based
on set~1 of \cite{Unintegr2}, where only the singular terms in the
gluon splitting function are included. Switching to 
set~2 of \cite{Unintegr2}, for which the full gluon splitting function
is used, results in a reduction in
the predicted cross section by up to 30 \%.
Set~2 gives the best description for different observables in another
recent dijet measurement covering a similar kinematic
region~\cite{Poeschl}.

%%%%%%%%%%%%%%%%%%%%%%%%%%%%%%%%%%%%%%%%%%%%%%%%%%%%%%%
%
%
\section{Summary}
\label{summary}
Triple differential dijet cross sections in $e^\pm p$
interactions are measured in the region of 
photon virtualities $2<Q^2<80\,{\GeV}^2$ and over a wide range
of inelasticities $0.1<y<0.85$. The data, covering the kinematic range
$\Etone>7\, \GeV$, $\Ettwo>5\, \GeV$ and pseudorapidities $-2.5<
\etaone , \etatwo <0$,
are compared with NLO and LO calculations,
with and without resolved photon contributions or
parton showers, as well as with a calculation based 
on $k_T$ factorisation and an unintegrated PDF of the proton.

A sizable and systematic excess of the data over the NLO
calculations of DISENT, which do not include a resolved
virtual photon contribution, is observed for $\qq < 10 \, \GeV^2$, 
small jet transverse energies, \Etmy\!,
and small $\xgmy$,
or equivalently, large jet pseudorapidities, $\etamy$.
The excess observed for
$\xgmy < 0.75$ decreases with increasing $Q^2$.

NLO QCD calculations  incorporating a resolved virtual
photon, as implemented in JETVIP,
bring the QCD predictions 
closer to the data, though there is still a deficit
at low $\xgmy$, especially for low $Q^2$. Unfortunately the JETVIP prediction
for the resolved part of the dijet cross section is 
sensitive to the choice of the slicing
parameter \ycut and must therefore be taken with caution.

The significant role of initial and final state QCD parton 
showers, which are not taken into account in the NLO QCD 
calculations, is demonstrated.  The inclusion
of QCD parton showers in the HERWIG LO Monte Carlo model 
leads to a considerable improvement
in the description, though a 
discrepancy remains in the region of high $\xgmy$.
The best agreement with the data is obtained when
both transversely and longitudinally polarised resolved virtual 
photons are included.

CASCADE, which is based on the CCFM evolution scheme, does not
involve the concept of virtual photon structure.
The CASCADE description of the data is best in the region 
of moderate $Q^2$ between 10 and 25\;$\GeV^2$.
The $Q^2$ dependence of the cross section
is less steep than in the data.

To conclude, the data show clear evidence for effects 
that go beyond 
the fixed-order NLO QCD calculations.
The importance of QCD parton showers and of the resolved \gammaL
contribution is demonstrated.

%%%%%%%%%%%%%%%%%%%%%%%%%%%%%%%%%%%%%%%%%%%%%%%%%%%%%%%
%
%
\section*{Acknowledgements}

We are grateful to the HERA machine group whose outstanding
efforts have made this experiment possible. 
We thank
the engineers and technicians for their work in constructing and
maintaining the H1 detector, our funding agencies for 
financial support, the
DESY technical staff for continual assistance
and the DESY directorate for support and for the
hospitality which they extend to the non-DESY 
members of the collaboration.

%%%%%%%%%%%%%%%%%%%%%%%%%%%%%%%%%%%%%%%%%%%%%%%%%%%%%%%%%
%
%

%%%%%%%%%%%%%%%%%%%%%%%%%%%%%%%%%%%%%%%%%%%%%%%%%%%%%%%%%
%
%%% DISENT ============================================

%%% JETVIP =============================================
%
%\begin{figure}\centering
%\epsfig{file=pict/qe3x.nif20.05.11.2003.eps,height=20cm}
%\caption{The triple differential dijet cross section \dsqex
%      from H1 data, compared with the predictions of
%      JETVIP NLO calculations corrected for hadronisation
%      effects.  The dashed line shows the NLO direct cross section,
%      the full line shows the sum of NLO direct and NLO resolved
%      contributions.  The theoretical uncertainty is not shown.}
%\label{qe3x.jet}
%\end{figure}
%
%%% Parton level =============================================

%%%% HERWIG and CASCADE ===============================

%%%%%%%%%%%%%%%%%%%%%%%%%%%%%%%%%%%%%%%%%%%%%%%%%%%%%%%%%
%
\begin{table}[hbt]
\begin{center}
%\small
\footnotesize
\begin{tabular}{ccccccc}
\hline
%$\qq (\gevsq)$  & $\Etmy (\GeV)$  &  \xg  & \sigone & \delstat & \delplus & \delminus \\
$\qq$  & $\Etmy $  &  \xg  & \dsqex & \delstat & \delplus & \delminus \\
$(\gevsq)$&$(\GeV)$&       &$(\pb/\GeV^3)$&$(\pb/\GeV^3)$&$(\pb/\GeV^3)$& \\
\hline
\qiaa&\eiaa&\xiaa&  94  &  1  & 20  &0.78\\
     &     &\xibb&  85.3&  1.0&  9.9&0.81\\
     &     &\xicc&  74.8&  0.9&  7.9&1.17\\
     &     &\xidd&  81  &  1  & 10  &1.04\\
     &\eibb&\xiaa&  10.0&  0.2&  1.8&0.87\\
     &     &\xibb&  13.6&  0.2&  2.0&0.88\\
     &     &\xicc&  17.6&  0.2&  2.1&1.03\\
     &     &\xidd&  35.6&  0.4&  3.4&1.01\\
     &\eicc&\xiaa&   0.033&  0.004&  0.010&0.81\\
     &     &\xibb&   0.141&  0.010&  0.029&0.93\\
     &     &\xicc&   0.239&  0.012&  0.045&0.94\\
     &     &\xidd&   0.90&  0.03&  0.15&0.98\\
\qibb&\eiaa&\xiaa&  26.9&  0.3&  5.1&0.80\\
     &     &\xibb&  29.5&  0.3&  3.2&0.82\\
     &     &\xicc&  26.2&  0.2&  2.8&1.19\\
     &     &\xidd&  40.9&  0.4&  4.8&1.05\\
     &\eibb&\xiaa&   3.20&  0.06&  0.60&0.86\\
     &     &\xibb&   5.03&  0.07&  0.66&0.90\\
     &     &\xicc&   6.31&  0.07&  0.84&1.02\\
     &     &\xidd&  16.3&  0.2&  1.5&1.02\\
     &\eicc&\xiaa&   0.0169&  0.0022&  0.0050&0.88\\
     &     &\xibb&   0.061&  0.004&  0.012&0.97\\
     &     &\xicc&   0.103&  0.006&  0.019&0.89\\
     &     &\xidd&   0.451&  0.014&  0.072&0.98\\
\qicc&\eiaa&\xiaa&   6.50&  0.09&  0.91&0.83\\
     &     &\xibb&   8.03&  0.08&  0.86&0.85\\
     &     &\xicc&   8.87&  0.08&  0.94&1.21\\
     &     &\xidd&  15.8&  0.1&  1.9&1.05\\
     &\eibb&\xiaa&   0.76&  0.02&  0.14&0.92\\
     &     &\xibb&   1.51&  0.02&  0.21&0.88\\
     &     &\xicc&   2.09&  0.03&  0.27&1.03\\
     &     &\xidd&   6.22&  0.06&  0.57&1.02\\
     &\eicc&\xiaa&   0.0048&  0.0007&  0.0016&1.05\\
     &     &\xibb&   0.0182&  0.0013&  0.0031&0.84\\
     &     &\xicc&   0.0403&  0.0019&  0.0079&0.96\\
     &     &\xidd&   0.155&  0.005&  0.028&0.98\\
\qidd&\eiaa&\xiaa&   0.93&  0.02&  0.16&0.86\\
     &     &\xibb&   1.30&  0.02&  0.13&0.89\\
     &     &\xicc&   1.65&  0.02&  0.27&1.26\\
     &     &\xidd&   3.97&  0.03&  0.46&1.06\\
     &\eibb&\xiaa&   0.165&  0.006&  0.027&0.89\\
     &     &\xibb&   0.330&  0.007&  0.045&0.91\\
     &     &\xicc&   0.451&  0.006&  0.057&1.06\\
     &     &\xidd&   1.75&  0.02&  0.17&1.02\\
     &\eicc&\xiaa&   0.00246&  0.00081&  0.00060&1.06\\
     &     &\xibb&   0.0052&  0.0005&  0.0015&0.88\\
     &     &\xicc&   0.0071&  0.0004&  0.0020&0.95\\
     &     &\xidd&   0.0534&  0.0015&  0.0097&0.98\\

\hline
\end{tabular}
\caption{Triple differential dijet cross section, \dsqex\!. 
    The cross section is given together with the 
    statistical and systematic uncertainties. The correction factors 
    for hadronisation effects applied to the NLO QCD predictions
    are also given.}
\label{tabqe3x}
\end{center}
\end{table}
%%%%%%%%%%%%%%%%%%%%%%%%%%%%%%%%%%%%%%%%%%%%%%%%%%%
\begin{table}[hbt]
\begin{center}
%\footnotesize
\scriptsize
%\tiny
\begin{tabular}{ccccccc}
\hline
%$\qq (\gevsq)$  & $y$  &  \etamy  & \sigtwo & \delstat & \delplus & \delminus \\
$\qq$     & $y$  &  \etamy  & \dsqyr & \delstat & \delplus & \delminus \\
$(\gevsq)$&      &          & $(\pb/\GeV^2)$&$(\pb/\GeV^2)$&$(\pb/\GeV^2)$& \\
\hline
\qiaa&\yiaa&\riaa& 870& 10&110&0.89\\
     &     &\ribb& 800& 10&110&1.01\\
     &     &\ricc& 643& 11& 98&1.03\\
     &     &\ridd& 480&  8& 66&0.99\\
     &     &\riee& 404&  8& 64&0.94\\
     &\yibb&\riaa& 531&  7& 48&1.07\\
     &     &\ribb& 362&  5& 38&1.04\\
     &     &\ricc& 297&  5& 42&0.97\\
     &     &\ridd& 234&  4& 37&0.94\\
     &     &\riee& 193&  4& 30&0.92\\
     &\yicc&\riaa& 224&  4& 21&1.03\\
     &     &\ribb& 168&  3& 17&0.95\\
     &     &\ricc& 130&  2& 15&0.91\\
     &     &\ridd& 123&  3& 23&0.91\\
     &     &\riee& 109&  3& 18&0.88\\
\qibb&\yiaa&\riaa& 377&  4& 44&0.90\\
     &     &\ribb& 363&  4& 41&1.02\\
     &     &\ricc& 266&  3& 30&1.04\\
     &     &\ridd& 192&  2& 24&0.99\\
     &     &\riee& 154&  2& 23&0.96\\
     &\yibb&\riaa& 209&  2& 18&1.10\\
     &     &\ribb& 137&  2& 13&1.06\\
     &     &\ricc&  97&  1& 11&1.00\\
     &     &\ridd&  75&  1& 10&0.96\\
     &     &\riee&  62.2&  1.0&  9.7&0.93\\
     &\yicc&\riaa&  80.7&  1.5&  6.8&1.05\\
     &     &\ribb&  55.3&  1.1&  5.6&0.98\\
     &     &\ricc&  41.4&  0.9&  4.2&0.94\\
     &     &\ridd&  37.6&  0.8&  5.6&0.90\\
     &     &\riee&  31.9&  0.8&  6.1&0.89\\
\qicc&\yiaa&\riaa& 133&  1& 16&0.89\\
     &     &\ribb& 132&  1& 14&1.04\\
     &     &\ricc&  94&  1& 11&1.05\\
     &     &\ridd&  63.7&  0.8&  8.0&1.04\\
     &     &\riee&  48.3&  0.7&  7.9&0.97\\
     &\yibb&\riaa&  71.2&  0.8&  6.3&1.11\\
     &     &\ribb&  45.8&  0.6&  4.3&1.09\\
     &     &\ricc&  30.8&  0.4&  3.0&1.01\\
     &     &\ridd&  23.1&  0.3&  2.7&0.97\\
     &     &\riee&  17.6&  0.3&  2.3&0.95\\
     &\yicc&\riaa&  25.0&  0.5&  2.0&1.06\\
     &     &\ribb&  16.4&  0.3&  1.5&0.99\\
     &     &\ricc&  12.1&  0.3&  1.4&0.94\\
     &     &\ridd&   8.4&  0.2&  1.3&0.93\\
     &     &\riee&   7.3&  0.2&  1.2&0.91\\
\qidd&\yiaa&\riaa&  28.0&  0.3&  4.1&0.90\\
     &     &\ribb&  29.5&  0.3&  3.2&1.04\\
     &     &\ricc&  19.8&  0.2&  2.6&1.08\\
     &     &\ridd&  13.8&  0.2&  2.3&1.05\\
     &     &\riee&   9.4&  0.2&  1.9&0.99\\
     &\yibb&\riaa&  16.8&  0.2&  1.5&1.13\\
     &     &\ribb&  11.1&  0.2&  1.0&1.11\\
     &     &\ricc&   7.31&  0.11&  0.70&1.03\\
     &     &\ridd&   4.69&  0.08&  0.51&1.02\\
     &     &\riee&   3.68&  0.08&  0.43&0.98\\
     &\yicc&\riaa&   6.43&  0.17&  0.52&1.09\\
     &     &\ribb&   4.32&  0.14&  0.39&1.00\\
     &     &\ricc&   2.78&  0.10&  0.28&0.98\\
     &     &\ridd&   1.95&  0.08&  0.26&0.96\\
     &     &\riee&   1.42&  0.06&  0.18&0.94\\

\hline
\end{tabular}
\caption{Triple differential dijet cross section, \dsqyr\!.
     See the caption of Table~\ref{tabqe3x} for further details.}
\label{tabqy1r5}
\end{center}
\end{table}
%%%%%%%%%%%%%%%%%%%%%%%%%%%%%%%%%%%%%%%%%%%%%%%%%%%
\begin{table}[hbt]
\begin{center}
%\footnotesize
\scriptsize
%\tiny
\begin{tabular}{ccccccc}
\hline
%$\qq (\gevsq)$  & \etamy & $\Etmy (\GeV)$ & \sigthree & \delstat & \delplus & \delminus \\
$\qq$  & \etamy & $\Etmy$ & \dsqre & \delstat & \delplus & \delminus \\
$(\gevsq)$&     &$(\GeV)$ &$(\pb/\GeV^3)$&$(\pb/\GeV^3)$&$(\pb/\GeV^3)$& \\
\hline
\qiaa&\ryaa&\eyaa&  28.6&  0.4&  2.5&1.04\\
     &     &\eybb&  54.9&  0.5&  5.5&0.99\\
     &     &\eycc&  16.6&  0.2&  1.8&0.97\\
     &     &\eydd&   2.78&  0.05&  0.42&0.94\\
     &     &\eyee&   0.276&  0.010&  0.076&0.92\\
%     &     &\ey&   0.00445&  0.00085&  0.00181&0.94\\
     &\rybb&\eyaa&  22.5&  0.4&  2.8&0.99\\
     &     &\eybb&  35.5&  0.4&  4.4&1.00\\
     &     &\eycc&  12.9&  0.2&  1.5&1.03\\
     &     &\eydd&   3.50&  0.08&  0.47&1.00\\
     &     &\eyee&   0.63&  0.02&  0.10&0.98\\
%     &     &\ey&   0.02442&  0.00257&  0.00552&0.89\\
     &\rycc&\eyaa&  19.4&  0.3&  2.7&0.89\\
     &     &\eybb&  24.6&  0.3&  3.8&0.95\\
     &     &\eycc&   8.1&  0.1&  1.0&1.00\\
     &     &\eydd&   2.07&  0.04&  0.29&1.00\\
     &     &\eyee&   0.450&  0.014&  0.072&0.99\\
%     &     &\ey&   0.03504&  0.00262&  0.00563&0.98\\
\qibb&\ryaa&\eyaa&  11.2&  0.1&  1.1&1.06\\
     &     &\eybb&  22.0&  0.2&  1.9&1.01\\
     &     &\eycc&   6.85&  0.06&  0.84&0.99\\
     &     &\eydd&   1.20&  0.02&  0.21&0.96\\
     &     &\eyee&   0.136&  0.005&  0.033&0.91\\
%     &     &\ey&   0.00136&  0.00018&  0.00080&0.90\\
     &\rybb&\eyaa&   7.60&  0.10&  0.79&1.00\\
     &     &\eybb&  13.0&  0.1&  1.5&1.04\\
     &     &\eycc&   5.22&  0.06&  0.50&1.03\\
     &     &\eydd&   1.59&  0.03&  0.21&1.01\\
     &     &\eyee&   0.298&  0.011&  0.052&0.97\\
%     &     &\ey&   0.01298&  0.00113&  0.00453&0.92\\
     &\rycc&\eyaa&   6.16&  0.06&  0.68&0.91\\
     &     &\eybb&   8.3&  0.1&  1.1&0.97\\
     &     &\eycc&   3.02&  0.03&  0.35&1.00\\
     &     &\eydd&   0.88&  0.01&  0.12&1.01\\
     &     &\eyee&   0.199&  0.005&  0.031&1.00\\
%     &     &\ey&   0.02307&  0.00207&  0.00367&0.98\\
\qicc&\ryaa&\eyaa&   3.64&  0.04&  0.32&1.06\\
     &     &\eybb&   7.41&  0.05&  0.64&1.02\\
     &     &\eycc&   2.41&  0.02&  0.29&0.99\\
     &     &\eydd&   0.445&  0.007&  0.071&0.94\\
     &     &\eyee&   0.045&  0.002&  0.011&0.93\\
%     &     &\ey&   0.00052&  0.00010&  0.00041&0.85\\
     &\rybb&\eyaa&   2.39&  0.03&  0.25&1.03\\
     &     &\eybb&   4.27&  0.04&  0.46&1.06\\
     &     &\eycc&   1.86&  0.02&  0.17&1.05\\
     &     &\eydd&   0.529&  0.010&  0.075&1.01\\
     &     &\eyee&   0.098&  0.003&  0.017&0.96\\
%     &     &\ey&   0.00459&  0.00044&  0.00123&0.97\\
     &\rycc&\eyaa&   1.80&  0.02&  0.20&0.94\\
     &     &\eybb&   2.41&  0.02&  0.30&0.98\\
     &     &\eycc&   0.98&  0.01&  0.11&1.03\\
     &     &\eydd&   0.313&  0.005&  0.040&1.02\\
     &     &\eyee&   0.078&  0.002&  0.011&1.00\\
%     &     &\ey&   0.00591&  0.00047&  0.00129&0.98\\
\qidd&\ryaa&\eyaa&   0.741&  0.009&  0.087&1.12\\
     &     &\eybb&   1.58&  0.01&  0.17&1.04\\
     &     &\eycc&   0.589&  0.006&  0.072&0.99\\
     &     &\eydd&   0.134&  0.002&  0.024&0.94\\
     &     &\eyee&   0.0143&  0.0006&  0.0046&0.92\\
%     &     &\ey&   0.00010&  0.00002&  0.00008&0.94\\
     &\rybb&\eyaa&   0.451&  0.007&  0.057&1.00\\
     &     &\eybb&   0.91&  0.01&  0.10&1.09\\
     &     &\eycc&   0.471&  0.006&  0.050&1.05\\
     &     &\eydd&   0.163&  0.003&  0.023&1.00\\
     &     &\eyee&   0.0324&  0.0012&  0.0068&0.99\\
%     &     &\ey&   0.00176&  0.00019&  0.00046&0.94\\
     &\rycc&\eyaa&   0.301&  0.004&  0.041&0.98\\
     &     &\eybb&   0.477&  0.005&  0.064&1.03\\
     &     &\eycc&   0.238&  0.003&  0.030&1.05\\
     &     &\eydd&   0.090&  0.002&  0.010&1.02\\
     &     &\eyee&   0.0229&  0.0007&  0.0042&1.00\\
%     &     &\ey&   0.00200&  0.00015&  0.00034&0.97\\

\hline
\end{tabular}
\caption{Triple differential dijet cross section, \dsqre\!.
     See the caption of Table~\ref{tabqe3x} for further details.}
\label{tabqre1}
\end{center}
\end{table}
%%%%%%%%%%%%%%%%%%%%%%%%%%%%%%%%%%%%%%%%%%%%%%%%%%%
\begin{table}[hbt]
\begin{center}
\footnotesize
%\scriptsize
%\tiny
\begin{tabular}{ccccccc}
\hline
%$\qq (\gevsq)$  & \xg & $y$ & \sigfour & \delstat & \delplus & \delminus \\
$\qq$  & \xg & $y$ & \dsqxy & \delstat & \delplus & \delminus \\
$(\gevsq)$&  &     &$(\pb/\GeV^2)$&$(\pb/\GeV^2)$&$(\pb/\GeV^2)$& \\
\hline
\qiaa&\xyaa&\yyaa& 586& 12&100&0.98\\
     &     &\yybb& 409&  9& 63&0.93\\
     &     &\yycc& 283&  7& 41&0.88\\
     &     &\yydd& 199&  5& 27&0.86\\
     &     &\yyee& 192&  8& 27&0.87\\
     &\xybb&\yyaa&1360& 31&200&0.96\\
     &     &\yybb& 653& 15& 63&1.11\\
     &     &\yycc& 289&  9& 25&1.15\\
     &     &\yydd& 164&  7& 14&1.13\\
     &     &\yyee&  89.6&  6.2&  7.3&1.12\\
\qibb&\xyaa&\yyaa& 198&  3& 27&0.99\\
     &     &\yybb& 126&  2& 17&0.95\\
     &     &\yycc&  92&  2& 12&0.90\\
     &     &\yydd&  66.9&  2.1&  9.1&0.88\\
     &     &\yyee&  53.6&  2.2&  7.1&0.88\\
     &\xybb&\yyaa& 744& 11& 74&0.97\\
     &     &\yybb& 291&  6& 30&1.12\\
     &     &\yycc& 138&  4& 12&1.17\\
     &     &\yydd&  69.0&  2.8&  5.5&1.15\\
     &     &\yyee&  34.3&  2.4&  2.7&1.15\\
\qicc&\xyaa&\yyaa&  58.8&  0.9&  8.6&1.04\\
     &     &\yybb&  37.8&  0.7&  4.6&0.97\\
     &     &\yycc&  25.0&  0.6&  3.1&0.91\\
     &     &\yydd&  17.5&  0.5&  2.2&0.91\\
     &     &\yyee&  13.4&  0.7&  1.5&0.88\\
     &\xybb&\yyaa& 294&  4& 30&0.97\\
     &     &\yybb& 112&  2& 12&1.13\\
     &     &\yycc&  49.6&  1.4&  4.0&1.17\\
     &     &\yydd&  24.0&  1.0&  1.8&1.14\\
     &     &\yyee&  12.35&  0.84&  0.91&1.14\\
\qidd&\xyaa&\yyaa&   9.5&  0.2&  1.8&1.07\\
     &     &\yybb&   6.98&  0.15&  0.85&1.04\\
     &     &\yycc&   5.00&  0.15&  0.51&0.97\\
     &     &\yydd&   3.32&  0.17&  0.36&0.93\\
     &     &\yyee&   2.25&  0.27&  0.22&0.92\\
     &\xybb&\yyaa&  72.5&  0.9&  8.6&0.98\\
     &     &\yybb&  31.4&  0.6&  2.9&1.12\\
     &     &\yycc&  15.0&  0.5&  1.3&1.14\\
     &     &\yydd&   8.99&  0.47&  0.88&1.16\\
     &     &\yyee&   4.01&  0.57&  0.85&1.13\\

\hline
\end{tabular}
\caption{Triple differential event cross section, \dsqxy\!.
     See the caption of Table~\ref{tabqe3x} for further details.}
\label{tabqx4y}
\end{center}
\end{table}

\end{document}